\documentclass[12pt]{article}
\usepackage{mainstyle}
\usepackage[backend=biber, maxnames=4, sorting=none, citestyle=numeric-comp, bibstyle=trad-unsrt]{biblatex}
\usepackage[a4paper,
            left=2cm,
            right=2cm,
            top=2cm,
            bottom=2cm]{geometry}

\usepackage{placeins}
\usepackage{subfig}
\usepackage{hhline}
\usepackage{siunitx}
\title{NNNN: Neural Networks for Newtonian Noise Mitigation\\
at the Einstein Telescope}

\author{Jan Kelleter$^a$\footnote{Corresponding author, email: jan.kelleter@rwth-aachen.de}, Patrick Schillings$^a$, Jonathan Kuckert$^a$, David Bertram$^{b}$\footnote{Now at the University of Cologne, Germany}, \\ Markus Bachlechner$^b$, Achim Stahl$^b$, Johannes Erdmann$^a$}

\date{\small $^a$ RWTH Aachen University, III. Physikalisches Institut A, Aachen, Germany \\
            $^b$ RWTH Aachen University, III. Physikalisches Institut B, Aachen, Germany \\ }

\newcommand*{\rind}{\ensuremath{\sqrt{r}_{\text{ind}}}}

\begin{document}

\maketitle

\begin{abstract}
\noindent

The gravitational effects of seismic waves, so-called Newtonian noise, will likely limit the low-frequency sensitivity of future ground-based gravitational wave detectors, such as the Einstein Telescope.
It has been proposed to mitigate this noise source by predicting it from measurements of the surrounding seismic displacement field using an array of seismometers.
In this paper, we investigate the Newtonian noise prediction abilities of neural networks based on synthetic data from such seismometer arrays and compare the results with the Wiener filter as benchmark.
We developed a simulation that generates density fluctuations of random plane waves and Gaussian wave packets, and that calculates the resulting Newtonian noise and displacement field.
We investigate the performance on approximately stationary wave fields and single dominating long- and short-term events.
For the first case, we observe comparable performance of neural networks and the Wiener filter with the networks performing slightly better.
For the second case, however, we find that convolutional neural networks and graph neural networks can outperform the Wiener filter by
factors of 15-80, depending on the frequency and the array configuration, and that they can reduce the corresponding Newtonian noise amplitude spectral density by factors of 10-30.
\\
\\
Keywords: Newtonian-noise mitigation, deep learning, Einstein Telescope
\end{abstract}

\section{Introduction}
\label{sec:Intro}
The first detection of a gravitational wave signal from a binary black hole system in 2015~\cite{firstdirectGWmeasurement} opened the field of gravitational wave astronomy.
With the current gravitational wave detectors LIGO~\cite{LIGOpaper}, Virgo~\cite{Virgopaper} and KAGRA~\cite{KAGRApaper, KAGRApaper2}, mergers of binary neutron stars~\cite{LIGOScientificBinaryNeutronStar} and mergers of neutron stars with black holes~\cite{LIGOScientificBHNNMerger} were also observed.
Proposed third-generation gravitational wave detectors, such as the Einstein Telescope,
promise increased sensitivity in measurements of binary coalescences as well as in searches for new sources of gravitational waves, such as supernova gravitational wave signals and stochastic gravitational-wave backgrounds~\cite{ETDesignReportUpdate}.

The extension of the sensitivity to smaller gravitational wave signals and to lower frequencies leads to new challenges in the reduction of a variety of noise sources for underground detectors~\cite{ETsensitivityCurve}. 
One of these noise sources is Newtonian noise (NN)~\cite{NewtonianNoiseOrigin}, expected to be the dominant noise source for the Einstein Telescope at frequencies below 15~Hz~\cite{et-sensitivity-curves}: Seismic waves cause density fluctuations in the rock around the mirrors or shift the cavern walls, creating a gravitational effect on the mirrors of the underground gravitational wave interferometer. 
To reduce NN, it has been proposed to measure the surrounding displacement field with an array of seismometers and then predict the NN in the instrument from these witness measurements~\cite{CellaPredictionOfNN}.
Wiener filters~\cite{WienerFilter} have been extensively used to test the prediction ability of such an array~\cite{terrestialGravityFluctuations,FrancescaSingleMirror, FrancescaJointMirrorOptimization, OurPaper, LIGO_NNreductionSimulation,ArrayOptimizationForVirgo,VirgoNewtonianNoiseCancellingSystem,VirgoNewtonianNoiseCancellingSystemPart2,ophardt2025silencing,rading2025distributed,vanBeveren:2023seq}.

To account for the non-stationary nature of NN, adaptive Wiener filters were developed~\cite{VirgoNewtonianNoiseCancellingSystemPart2}.
In Ref.~\cite{vanBeveren:2023seq}, however, it was shown that deep neural networks have the potential to outperform Wiener filters for the prediction of NN.
The study in Ref.~\cite{vanBeveren:2023seq}, combined signals from a surface array of geophones with signals from an array of seismometers and geophones in the tunnels of the Einstein Telescope.
These were then used as input to a neural network with fully-connected layers, also known as a multilayer perceptron (MLP).

In this paper, we investigate convolutional (CNN) and graph neural networks (GNN) for NN predictions and compare them to the Wiener filter.
These network architectures are able to directly exploit the local spatial structures of the input data.
We use a simplified 3D simulation of density fluctuations in a homogeneous medium that cause ground displacement and numerically determine the corresponding NN. We validate this simulation against analytical calculations. Both the Wiener filter and the neural network are trained to predict NN at a single test mass from the displacements measured in an underground array of seismometers.
The seismometer positions are not limited to the telescope's tunnels, as studies on the optimization of grids for seismometer arrays suggest that more general seismometer positions are needed to efficiently suppress NN~\cite{FrancescaSingleMirror, FrancescaJointMirrorOptimization, OurPaper, ArrayOptimizationForVirgo,ophardt2025silencing}.
While 3D CNNs are a natural choice for regular seismometer array grids, GNNs are an alternative for optimized irregular grids.

Section~\ref{sec:simulation} introduces the simulation and metrics used for the prediction of NN. Section~\ref{sec:WF} describes the Wiener filter as the benchmark method and presents the validation of the simulation against analytical calculations. In Section~\ref{sec:NN}, we introduce the neural networks and describe the training process. The results are presented in Section~\ref{sec:results}. Finally, a summary and an outlook are given in Section~\ref{sec:conclusions}.

\section{Newtonian noise from a toy simulation of seismic waves}
\label{sec:simulation}
For an underground detector, such as the Einstein Telescope, it has been suggested that out of the different types of seismic waves body waves produce the dominating contribution to NN~\cite{SiteSelectionCriteria}.
There are two kinds of body waves: P-waves and S-waves. The former, also known as primary waves or pressure waves, are longitudinal sound-like waves that cause density fluctuations in homogeneous rock. They can also shift interfaces of different density to produce fluctuating gravity fields. S-waves, also known as secondary or shear waves, shift rock perpendicular to their direction of movement and can only cause NN via the displacement of interfaces. This is especially relevant at the walls of the detector caverns.

To simulate\footnote{The code can be found at \url{https://github.com/lc316353/Newtonian-Noise-Simulation}.} P-waves, we start from density fluctuations $\delta \rho$ in homogeneous rock. They are modeled as monochromatic plane waves or Gaussian wave packets on a 3D grid $\vec{x}$ with $N_x$ points and a spacing of $\Delta x$. The size of the simulation cube in any direction is denoted by $2L$.
Each generated time series is evaluated at $N_t$ points starting from $t=0$ to $t_{\max}$.

Similar to Ref.~\cite{LIGO_NNreductionSimulation}, the packets are given by
\begin{equation}
    \delta\rho(\vec x,t)/\rho=A e^{2\pi^2\sigma_f^2 t'^2}\sin\bracket{2\pi f_0  t'+\phi},
\end{equation}
where $\rho$ is the constant density of the homogeneous rock, $A$ the amplitude, $\sigma_f$ the width of the packet and $f_0$ the base frequency, while $\phi$ is the phase (set to 0 to ensure that the wave packet does not transport mass, i.e., the volume integral over $\delta \rho$ vanishes); $t'$ is given by
\begin{equation}
    t'= \bracket{\vec x \cdot \vec e_\text{wave}-r_0}/c_P-\bracket{t-t_0},
\end{equation}
where the unit vector $\vec e_\text{wave}$ is the propagation direction of the wave, defined by the polar angle $\varphi$ and the azimuthal angle $\vartheta$, $r_0$ is the distance of the wave to the origin at start time $t_0$, and $c_P$ is the sound velocity in rock. 
When considering monochromatic plane waves, the frequency width $\sigma_f$ is simply set to~0.
The simple propagation of wave packets simulated here, implies that more complex effects, such as reflections and mode conversion of seismic waves, are not considered.

To numerically calculate the NN force along the mirror axis (which is $\vec e_M$) of a mirror at position $\vec x_M$, the density fluctuations need to be integrated:
\begin{equation}
    \frac{F_M(t)}{\rho}=GM\int \dd \vec x\, \frac{(\vec x-\vec x_M)\cdot \vec e_M}{\abs{\vec x-\vec x_M}^3}\frac{\delta\rho(\vec{x},t)}{\rho} \, K(\vec x) , 
\end{equation}
where $G$ is the gravitational constant, $M$ is the mass of the mirror and $K(\vec x)$ is a kernel function. The kernel function has value 0 for points outside the designated integration which allows, for example, to cut the medium above the surface of the Earth. In our case, the kernel creates a small cavern with radius $r_\text{cavern}$ around the mirror and restricts the outer integration borders to a sphere of radius $L$ for the straight-forward comparison with analytical calculations. 

The shift of the cavern wall due to P-waves is considered using the analytical result for the approximation of a small spherical cavern~\cite{terrestialGravityFluctuations}:
\begin{equation}
    F_\text{cavern}(t)=-\frac{4\pi}{3}GM\rho\, \vec \xi(\vec x_M,t)\cdot \vec e_M,
\end{equation}
where $\vec \xi(\vec x_M,t)$ is the displacement field at the mirror position. It is possible to exchange this expression for different cavern shapes.

Denoting the fraction of P-waves by $p$, each wave is treated as an S-wave with a probability of $1-p$. In this case, it does not have a bulk NN contribution and moves at speed $c_S$. The effect of the cavern wall is perpendicular to the direction of movement with a random polarization angle $\alpha$.

The displacement $\vec\xi$ at arbitrarily placed seismometers and at the mirror position(s) is calculated from the continuity equation assuming $\delta\rho\ll\rho$:
\begin{equation}
    \frac{\delta\rho}{\rho}=-\vec\nabla\cdot\vec\xi.
\end{equation}
To reduce this problem to one dimension, each density fluctuation is integrated along its direction of movement, shifted to the seismometer position, rotated accordingly and projected onto the three spatial axes. Seismometer noise can be added afterwards.

\begin{figure}
    \centering
    \includegraphics[width=0.99\linewidth]{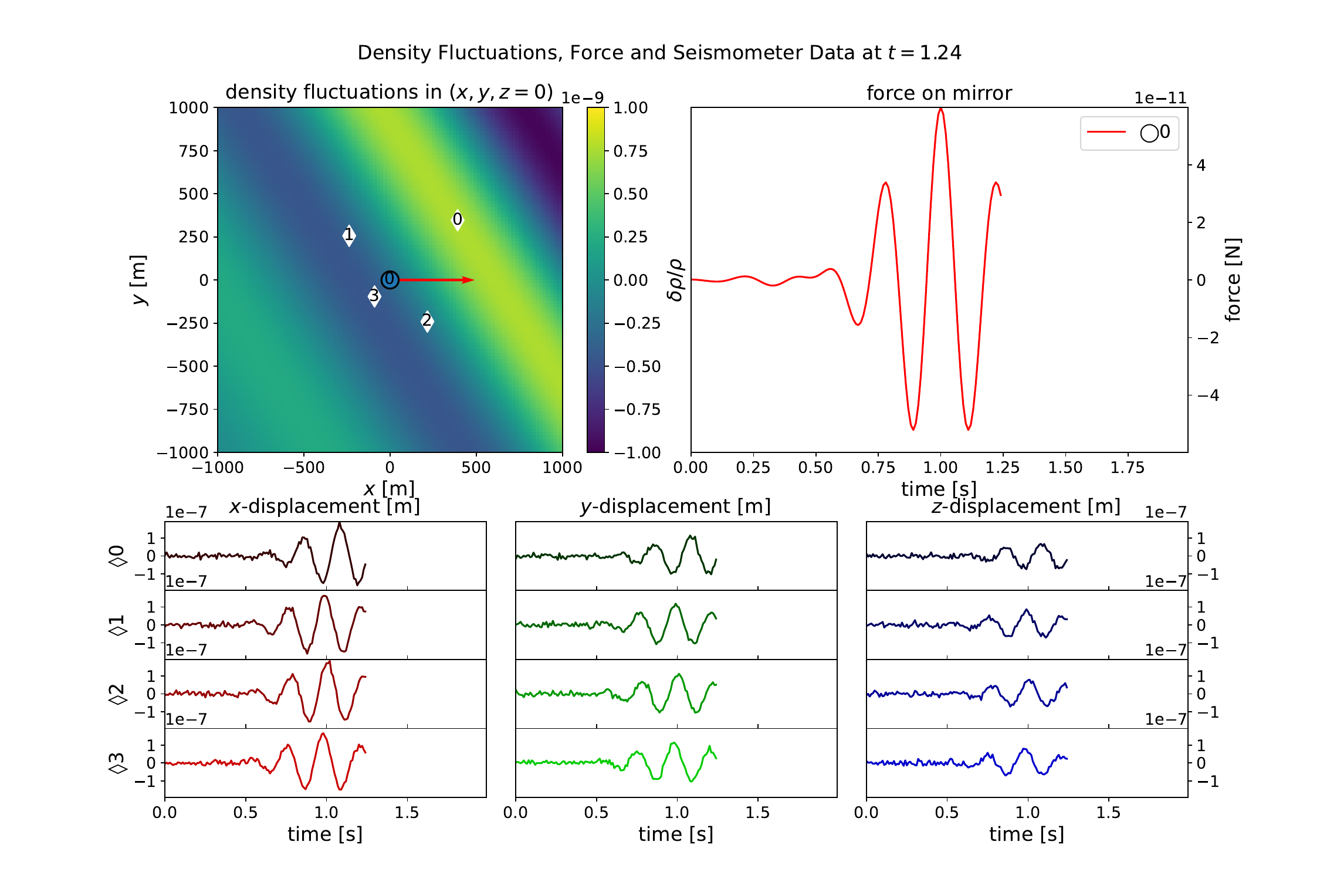}
    \caption{Display of a single data sample for the case of a Gaussian wave packet. In the top left, the density fluctuations are shown in two dimensions for $z=0$, i.e., in the plane that contains the mirror (blue circle). The fluctuations lead to a force on the mirror (red arrow). The time series of the force can be seen on the top right. The seismometers (white diamonds) record a displacement time series in 3 spatial channels each (lower plots).}
    \label{fig:simulationDisplay}
\end{figure}

In the end, the simulation produces a dataset containing $N$ wave events, each containing time series of the $x$-, $y$- and $z$-displacement of $N_S$ seismometers at given positions, and the time series of the NN force at $N_M$ mirrors. Here, we limit the studies to the movement of a single mirror along the $x$-direction. A visualization of a single wave packet event is shown in Fig.~\ref{fig:simulationDisplay}.
The values used for the simulation parameters used in this study are given in Table~\ref{tab:simulationParameters}.

\renewcommand{\arraystretch}{1}
\begin{table}
    \centering
    \begin{tabular}{|l|c|c|}
    \hline
       Simulation length  & $L$ & \SI{6\,000}{m}\\
       Number of grid points  & $N_x$ & $100$\\
       Simulation time  & $t_{\max}$ & \SI{2}{s}\\
       Number of time steps  & $N_t$ & $100$\\
       \hline
       Mirror mass & $M$ & \SI{211}{kg} \\
       Cavern radius & $r_\text{cavern}$ & \SI{5}{m}\\
       Sound velocity of P-waves in rock  & $c_P$ & $\SI{6\,000}{\meter\per\s}$\\
       Sound velocity of S-waves in rock  & $c_S$ & $\SI{4\,000}{\meter\per\s}$\\
       P-wave fraction & $p$ & $0.2$ \\
       \hline
       Reference time & $t_0$ & $[\SI{0.1}{s},\SI{1.9}{s})$\\
       Distance to origin at $t_0$ & $r_0$ & $\SI{-6\,000}{m}$\\
       Incident polar angle  & $\varphi$ & $[0,2\pi)$\\
       Incident azimuthal angle  & $\cos(\vartheta)$ & $[0,1)$\\
       S-wave polarization & $\alpha$ & $[0,2\pi)$\\
       Amplitude & $A$ & $[\SI{1e-9}{m},\SI{2e-9}{m})$\\
       Phase & $\phi$ & 0\\
       Frequency & $f_0$ & $[\SI{1}{Hz},\SI{10}{Hz})$\\
       Frequency width (= 0 for plane waves) & $\sigma_f$ & $[\SI{.5}{Hz},\SI{2}{Hz})$\\
       \hline
    \end{tabular}
    \caption{Values for the simulation parameters (domain, experimental setup and wave parameters). When a range is given, the parameter is sampled from a uniform distribution.}
    \label{tab:simulationParameters}
\end{table}

\section{Wiener filter and residual}
\label{sec:WF}
The Wiener filter (WF)~\cite{WienerFilter} estimates a signal based on knowledge of the correlations between the signal and the data, i.e., the seismometer measurements in this study. It can be applied in the time domain and in the frequency domain. 
The WF $\vec W_F$ operates on the data vector $\vec d$ (with one dimension per data point) as
\begin{equation}
    \tilde{s}_k=\vec W_{F,k}\cdot\vec d,
\end{equation}
where $\tilde{s}_k$ denotes a prediction of the mirror force $F_M(f_k)$ and $\vec d=(\xi_{1,x}(f_0),\xi_{1,y}(f_0),...,\xi_{N_S,z}(f_{N_f}))=(\vec{\xi}(f_0),...,\vec\xi(f_{N_f}))$ for each frequency bin $f_k$ in the frequency domain. The vector $\vec \xi(f_k)$ denotes the list of all $3N_S$ seismometer channels ($N_S$ seismometers that measure along the $x$-, $y$- and $z$-direction).
Alternatively, the WF can be expressed in the time domain by transforming $F_M(f_k) \mapsto F_M(t_k)$ and $\vec{\xi}(f_k) \mapsto \vec{\xi}(t_k)$.
In frequency space, a signal in one frequency bin does not depend on the data in any other frequency bin, making it possible to reduce the dimensions by changing $\vec d\rightarrow\vec \xi(f_k)$.

The WF is given by 
\begin{equation}
    \vec W_{F,k}=\EW{\vec d^* \vec d}^{-1}\cdot \EW{\vec d^* s_k},
\end{equation}
where $\cdot^*$ denotes the complex conjugate and both $\vec d$ and $s$ is the mirror force $F_M$ in either the time or frequency domain. The expectation value $\EW{\cdot}$ is the mean over the elements of the dataset. The matrix $\EW{\vec d^*\vec d}$ is called the data cross power spectral density (CPSD) in frequency space. 

The residual is defined in the frequency domain as 
\begin{equation}
    R=\frac{\EW{\abs{s-\vec W_F\cdot\vec{d}}^2}}{\EW{\abs{s}^2}}\, ,
    \label{eq:SimulatedResidual}
\end{equation}
and it is a function of frequency. The expectation value is over a test dataset.
It evaluates to 
\begin{equation}
    R=\frac{\EW{\vec d s^*}\cdot\EW{\vec d^*\vec d}^{-1}\cdot\EW{\vec d^*s}}{\EW{s^*s}}
    \label{eq:TheoreticalResidual}
\end{equation}
when inserting the definition of the WF.

\begin{figure}[h]
    \centering
    \includegraphics[width=0.5\linewidth]{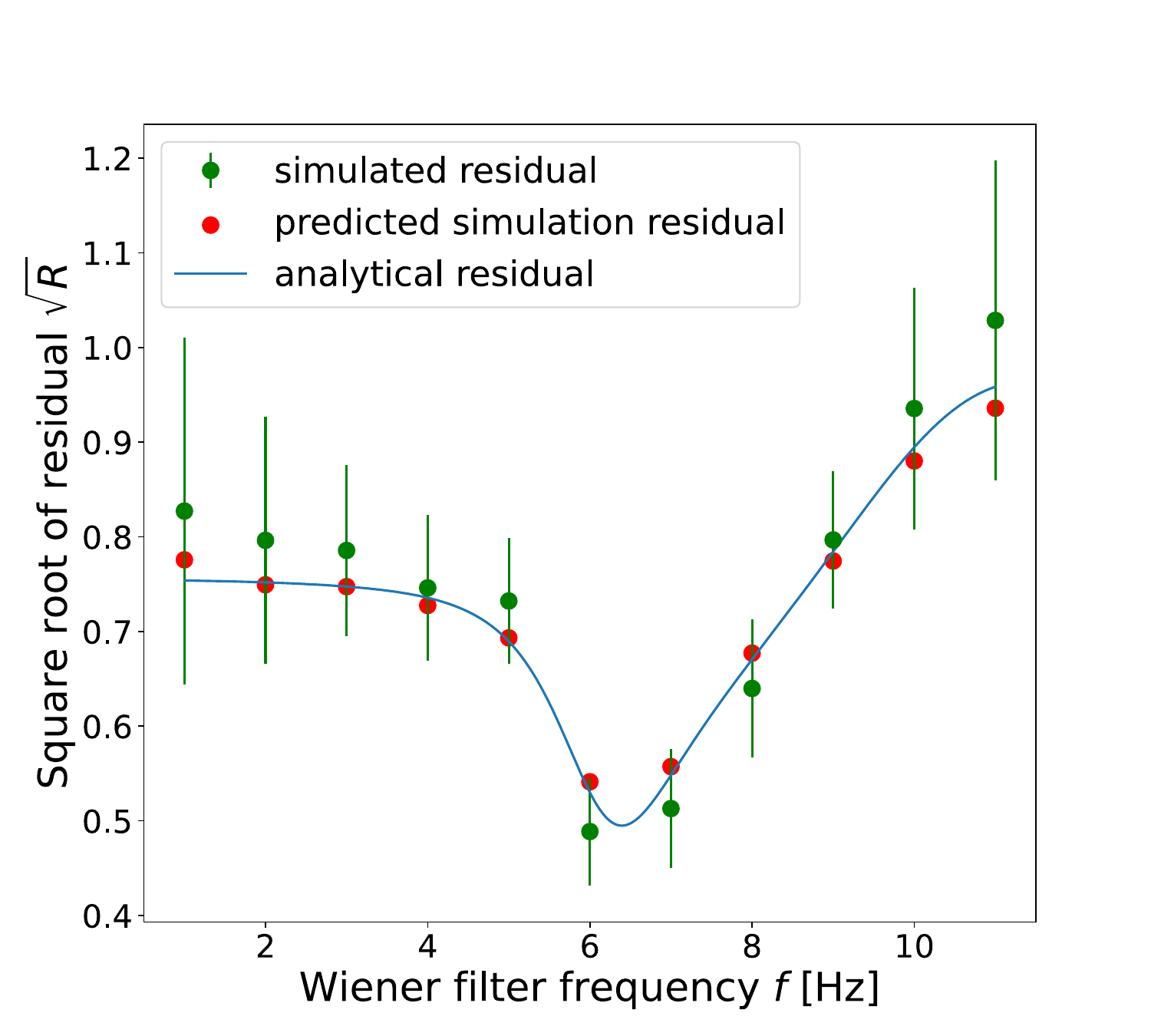}
    \caption{Comparison between the analytical prediction of Ref.~\cite{FrancescaSingleMirror} (blue) and our simulation (red and green). The square root of the residual $\sqrt{R}$ is plotted against the frequency $f$. The WF was calculated from the CPSDs of $9\,000$ wave events. With these, the theoretically predicted residual can be calculated from Eq.~\eqref{eq:TheoreticalResidual} (red). We apply the WF to a test set of $1\,000$ more wave events and calculate the residual by its definition in Eq.~\eqref{eq:SimulatedResidual} (green).}
    \label{fig:simulationCheckResidual}
\end{figure}

In order to cross-check the simulation, several tests were performed. In Fig.~\ref{fig:simulationCheckResidual}, we show one of these tests. In this case, 8 seismometers are placed in a regular cube with a side length of \SI{400}{m} and the mirror at its center. We do not add any noise for this cross-check study. We numerically determine the WF in the frequency domain on $9\,000$ isotropic, Gaussian wave packet events with random frequency widths and frequencies. 
We then calculate the square root of the residual $\sqrt{R}$ as the average over $1\,000$ events according to Eq.~\eqref{eq:SimulatedResidual}. The result can be seen as green dots with error bars showing the propagation of uncertainty of the error on the mean.
The red dots are calculated from Eq.~\eqref{eq:TheoreticalResidual}
from the resulting WF, i.e., without any test events.
One expects this to be the result of the evaluation on an infinitely large test dataset under the condition that the WF was calculated to sufficient numerical precision.
As the red dots agree with the green dots within their errors, we conclude that $9\,000$ events are sufficient for the determination of the WF.
The analytical expectation for this benchmark is shown as a solid blue line~\cite{FrancescaSingleMirror}. It agrees very well with the simulated residual, which serves as a validation for the numerical simulation.

\begin{figure}[h]
\makebox[\textwidth]{
    \centering    \subfloat{\includegraphics[width=.5\linewidth]{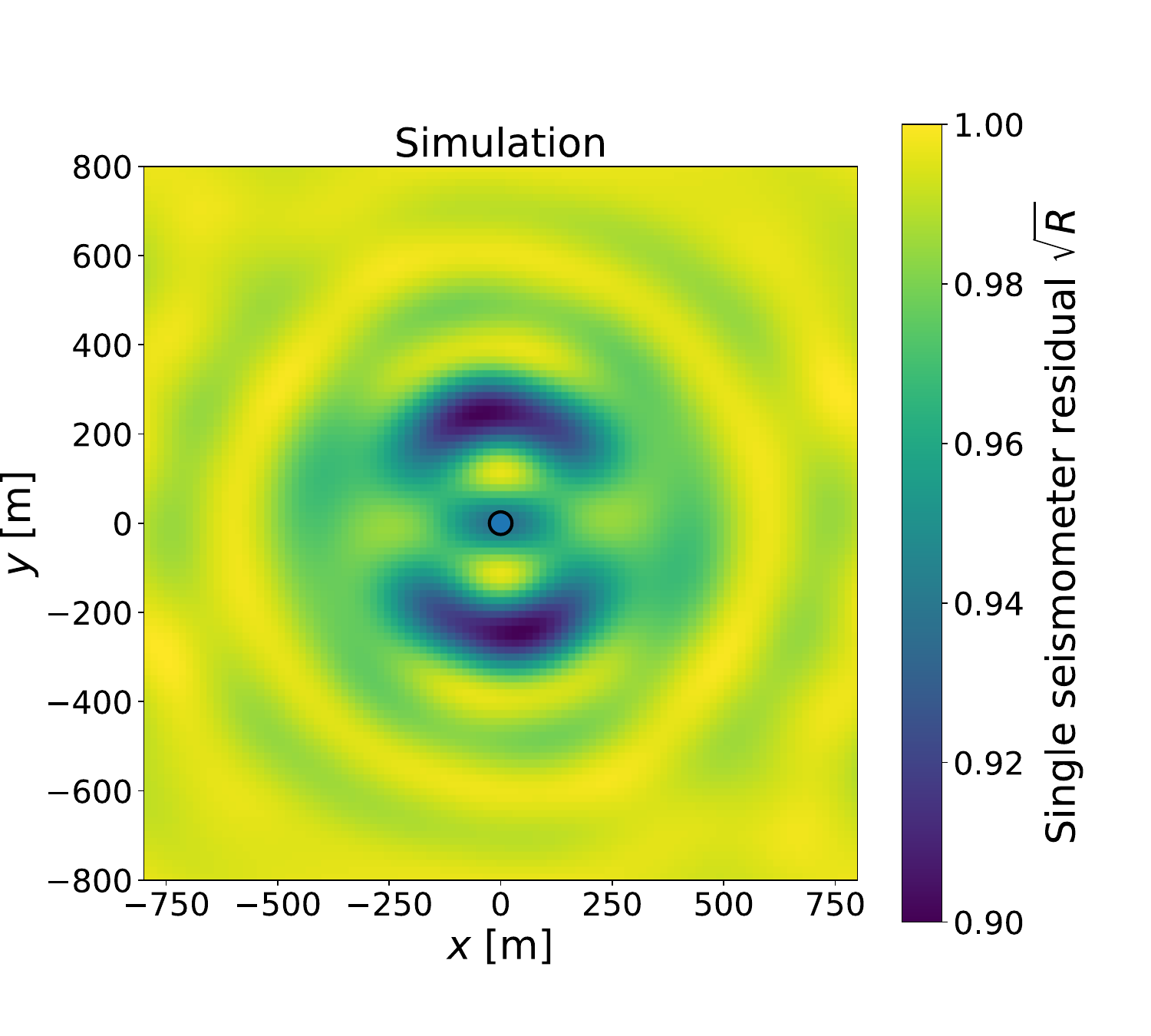}}
    \subfloat{\includegraphics[width=.5\linewidth]{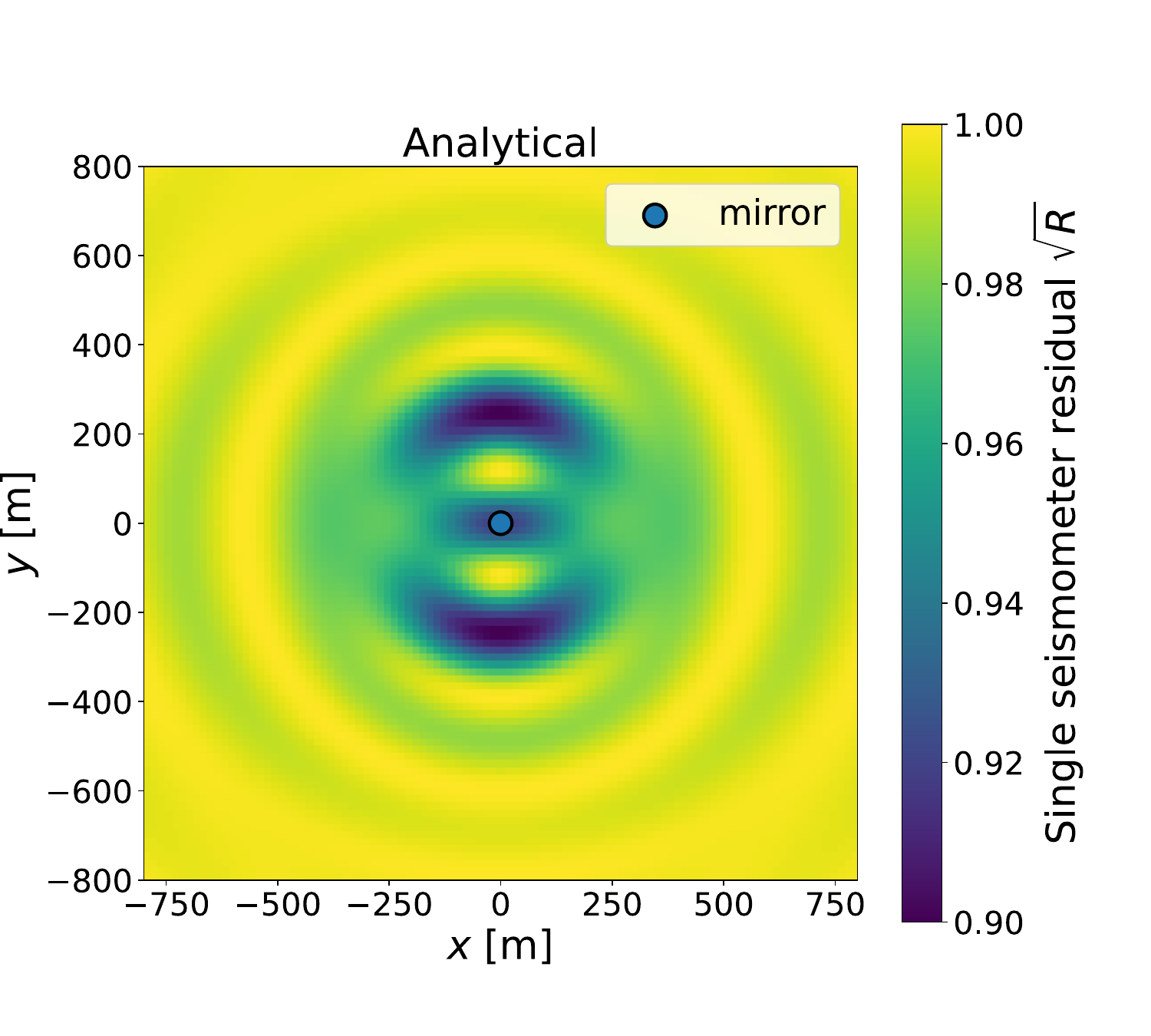}}
    }
    \caption{The square root of the residual for a single seismometer in the $x$-$y$-plane as a function of position for the simulation~(left) and the analytical calculation~(right) for $f=\SI{10}{\Hz}$ and $p=0.2$. For the simulation, $90\,000$ Gaussian wave packets were used to numerically determine the WF, and $10\,000$ wave packets were used as a test dataset.}
    \label{fig:simulationCheckGrid}
\end{figure}

Another cross-check is the NN mitigation performance of a single seismometer placed at different positions around the mirror. Figure~\ref{fig:simulationCheckGrid} shows the square root of the residual as a function of the seismometer position in the $x$-$y$-plane for the simulation and for the analytical calculation. Here, $90\,000$ wave packets are used to numerically determine the WF at 10~Hz, and $10,000$ wave packets to evaluate it. The simulation and the analytical calculation agree very well.

These tests validate the simulation, which we use to create datasets for the training and evaluation of WFs and neural networks in the following.
We note that the residuals in the presented cross-check studies agree with the analytical solutions, even though we have used wave packets instead of monochromatic waves.

\section{Neural network architecture and training}
\label{sec:NN}

\begin{figure}[t]
    \centering
    \includegraphics[width=0.99\linewidth]{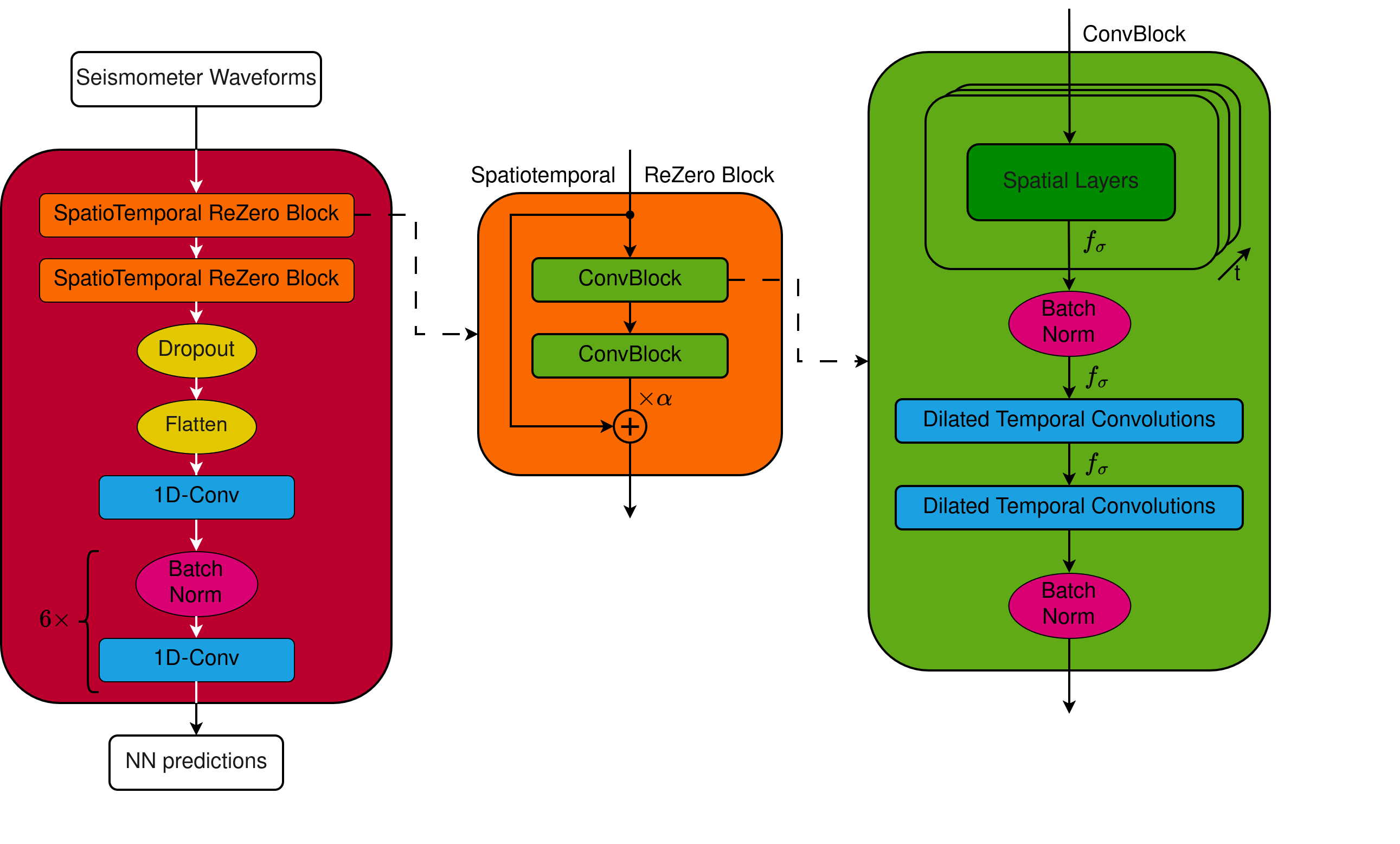}
    \caption{Schematic of the network architecture, with increasing granularity from left to right. The separation of spatial and temporal convolutions allows for a high amount of flexibility. For regular (irregular) grids, CNN layers (graph attention layers) were used as spatial layers.}
    \label{fig:net_setup}
\end{figure}

The neural network architecture\footnote{The code can be found at \url{https://gitlab.git.nrw/jan.kelleter/nnnn.git}.} employed is based on a spatiotemporal approach, in which separate convolutions over the spatial and temporal dimensions are performed. An overview of the architecture is shown in Fig.~\ref{fig:net_setup}. A more detailed list of hyperparameters used for network and training can be found in Table~\ref{tab:net_params}. The central building block of the architecture is the ConvBlock, in which the spatial and temporal convolutions take place. Temporal convolutions are dilated by factors 2-8, widening with each temporal convolution, to allow for the detection of features separated further in time. Two ConvBlocks are wrapped in a ReZero block~\cite{ReZero}, in which the output is weighted with a trainable parameter $\alpha$ and added to the input. After two `SpatioTemporal ReZero blocks', the output is flattened and a set of 1D convolutions is performed to create the time series prediction for the NN.

\begin{table}[h!]
    \centering
    \begin{tabular}{|l|c c|}
    \hline
        \textbf{Network Parameters} & CNN & Graph\thickspace \\
        \hline
        Number of layers & 2 & 8 \\
        Spatial kernel size and channels & $3\times3\times3\times64$ & N.A.\\
        Spatial attention heads and channels & N.A. & $4\times 64$ \\
        Spatial padding & Zero padding & None \\
        Dropout rate & \multicolumn{2}{c|}{0.4} \\
        Temporal kernel size and channels & \multicolumn{2}{c|}{$3 \times 64$}\\
        Final 1D-Conv kernel size and channels & \multicolumn{2}{c|}{$3\times 64$} \\
        Activation function & \multicolumn{2}{c|}{LeakyRelu, $\beta=0.5$} \\
        \hhline{|= = =|}
        \multicolumn{3}{|l|}{\textbf{Training Parameters}} \tabularnewline
        \hline
        Batch size & \multicolumn{2}{c|}{$256$}\\
        Initial learning rate & \multicolumn{2}{c|}{$10^{-3}$}\\
        Learning rate scheduler factor & \multicolumn{2}{c|}{$0.1$}\\
        Optimizer & \multicolumn{2}{c|}{Nadam}\\
        Loss & \multicolumn{2}{c|}{Mean absolute error}\\
        \hline
    \end{tabular}
    \caption{Neural network hyperparameters.}
    \label{tab:net_params}
\end{table}

Separating spatial and temporal convolutions allows for more flexibility, in particular the use of different operations on spatial and temporal dimensions. More memory efficient CNNs~\cite{CNNs} were used in the spatial part for larger seismometer configurations, while Graph Attention Layers (GATs)~\cite{GATs} were used to test the benefits of irregular seismometer placements. Graph neural networks are provided information about the arrangement of nodes via an adjacency matrix, where we create an adjacency matrix which holds the inverse distance between each seismometer pair in its entries and 1 for the self-adjacencies.

We use three different datasets. One containing a single Gaussian wave packet per event, one containing a single plane wave per event, and one containing the sums of 10 individually sampled plane waves per event as an approximate stationary case. We use 1 million samples with a train:validation:test split of 80:10:10. The data were simulated as described in Sections~\ref{sec:simulation} and~\ref{sec:WF} in the time domain. The data were normalized to zero mean and unit variance across the entire dataset and white noise with a {signal-to-noise ratio} of 15 with respect to the mean of the dataset was added to simulate seismometer noise. For comparison, time domain WFs were also calculated and evaluated on the training and test data, respectively.

We perform tests for different seismometer and network configurations. A baseline case of 8 seismometers arranged in a cube with side length $\SI{400}{\m}$, centered around a mirror serves to compare the networks against each other. A second, larger grid, consisting of 32 seismometers arranged in a $4\times4\times2$ cuboid with side lengths $\SI{800}{\m}\times\SI{800}{\m}\times\SI{400}{\m}$ was used to test the effects of an increased number of seismometers. Due to technical constraints, this larger grid was, at the moment, only processed with CNNs. Finally, an irregular grid of 8 seismometers was used to test the benefits of seismometer position optimization. The positions were optimized as described in Ref.~\cite{FrancescaSingleMirror} with Particle Swarm Optimization~\cite{PSOAlgorithm}. This configuration was only evaluated with the GNNs, since CNNs rely on a regular grid of sensors.

\section{Results}
\label{sec:results}

\subsection{The Stationary Case}

\begin{figure}
    \centering
    \includegraphics[width=0.6\linewidth]{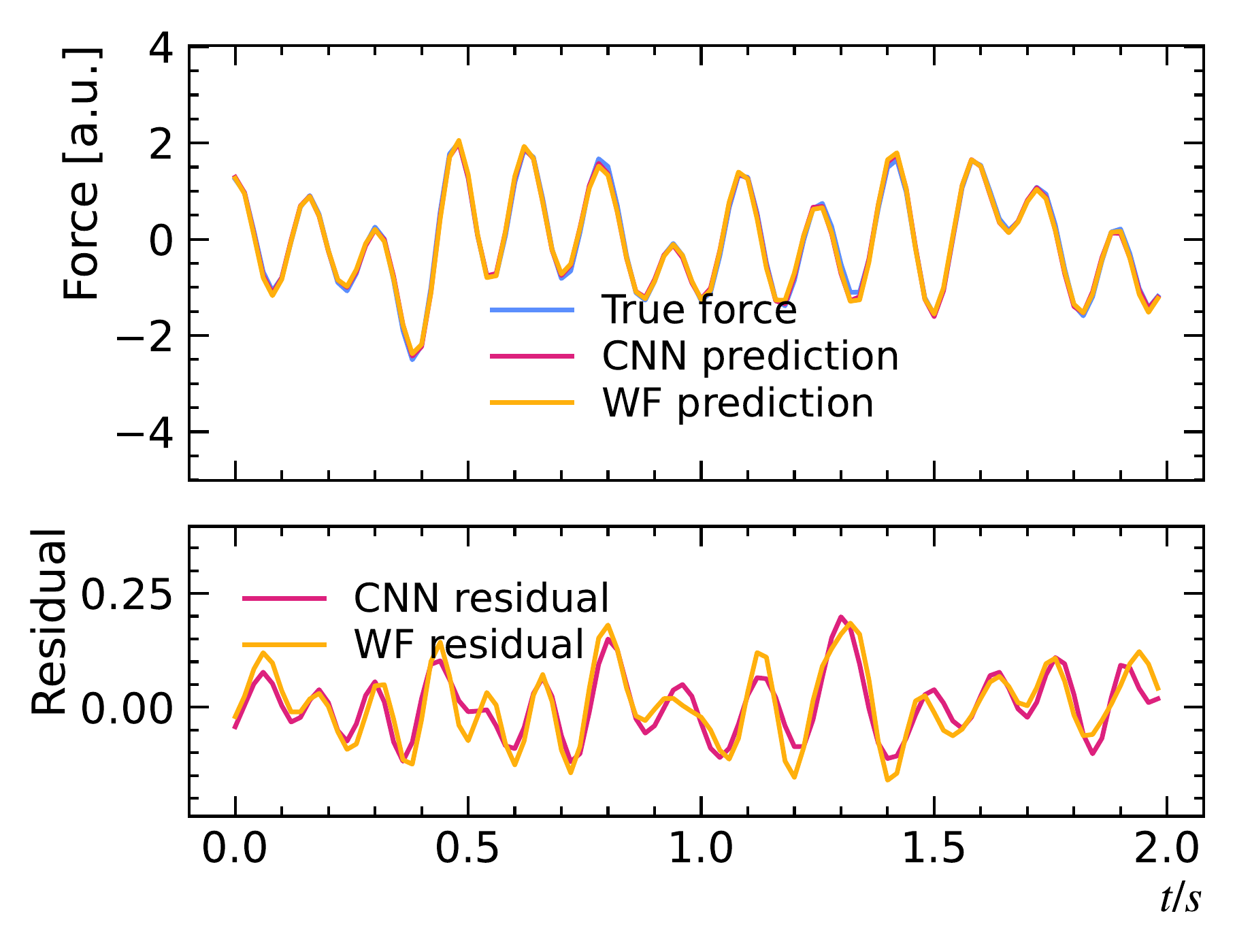}
    \caption{Single event prediction of the WF and CNN using a regular {8-seismometer} grid, and the residual signal after subtraction. The predictions follow the simulated truth closely, the maximum amplitude is reduced by an order of magnitude for both mitigation approaches.}
    \label{fig:timeseries}
\end{figure}

The first case considered is that of 10 plane waves overlaid to approximate a stationary wave field. For illustration, we plot the prediction of a WF and a CNN trained using the regular 8-seismometer grid against the true NN signal for a single example timeseries in Fig.~\ref{fig:timeseries}. The top plot shows the simulated force at the mirror due to NN, as well as the predictions of the WF and CNN. The bottom shows the residual signals of WF and CNN after subtraction. Both WF and CNN are able to match the simulated force closely and the residuals show a reduction of noise amplitude by about an order of magnitude.

\begin{figure}
    \makebox[\textwidth]{
        \subfloat[]{\includegraphics[width=0.53\linewidth]{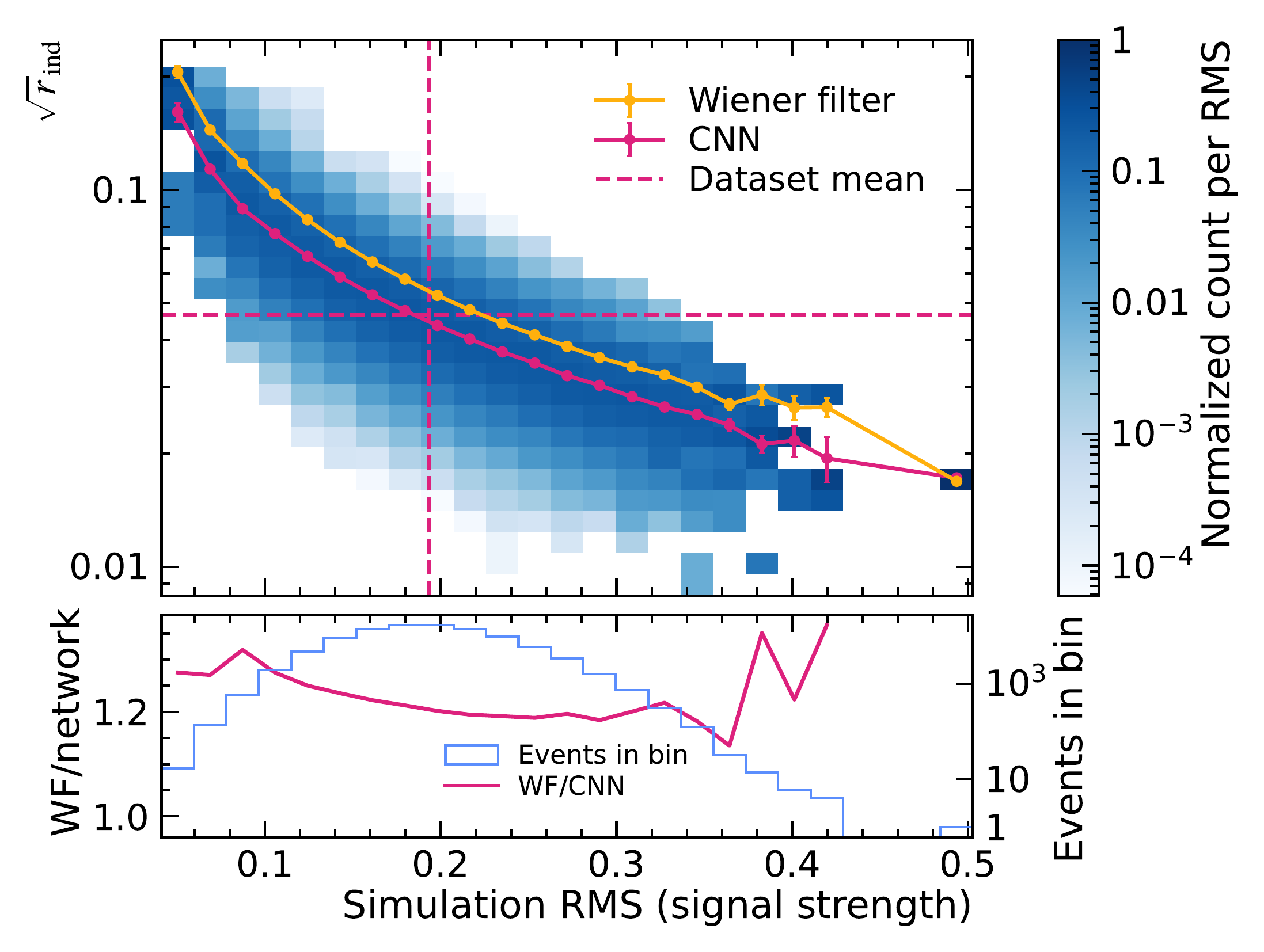}\label{fig:CNN_multi_r}}
        \subfloat[]{\includegraphics[width=0.47\linewidth]{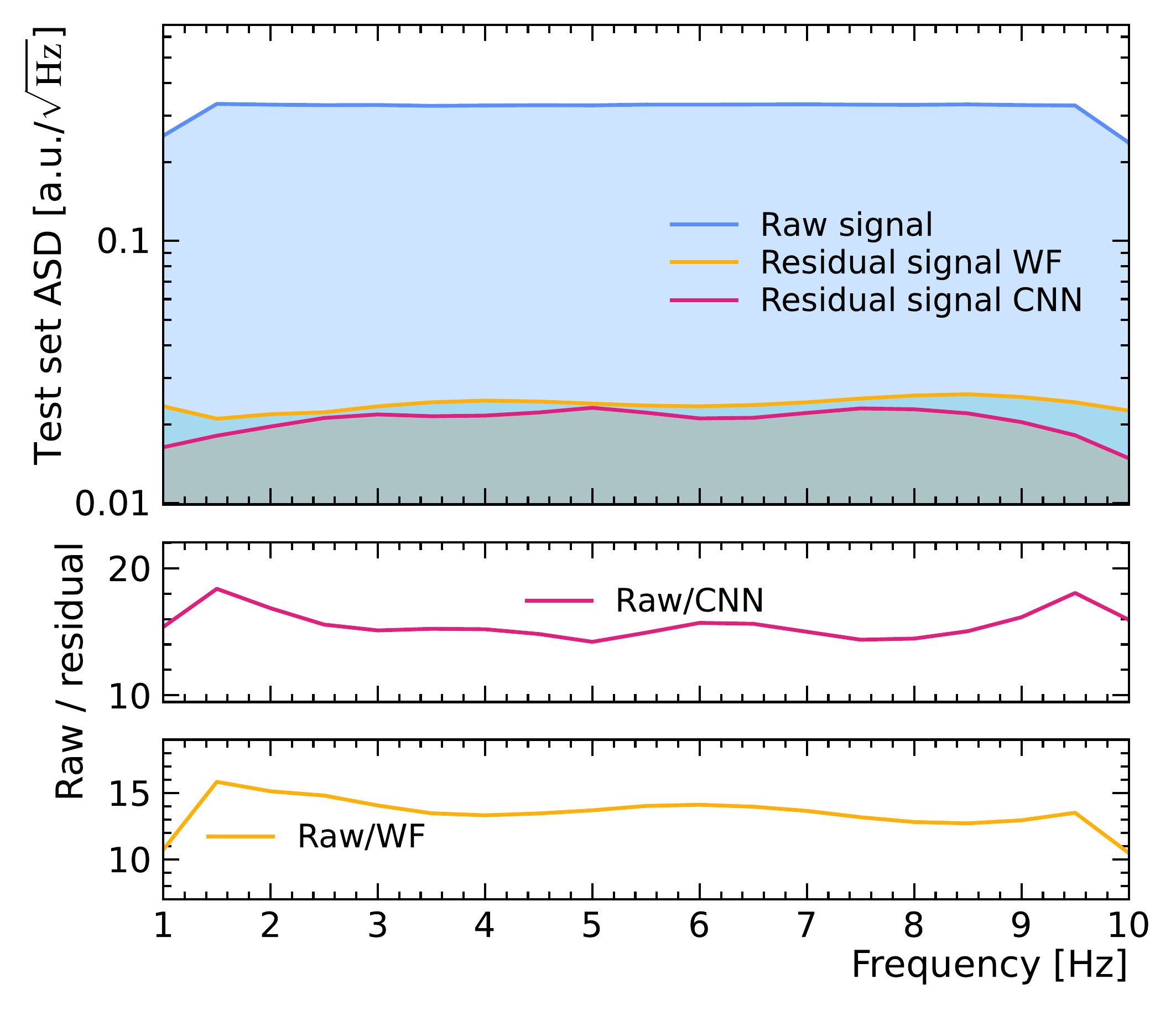}\label{fig:CNN_multi_ASD}}
    }
    \caption{Results of training a CNN and WF on the regular {8-seismometer} grid for 10 overlaid plane waves approximating a stationary wave field case: (a) Residual amplitudes as a function of initial event strength. Values below one denote mitigation, while values above one denote amplification. The average residual across the entire dataset is below $5\%$. Both CNN and WF perform similarly with factors 1.1 to 1.4 between their residuals, where the performance of the CNNs is slightly better. (b) ASDs of raw, WF- and CNN-mitigated events. Both have mitigation factors between 10 and 20 across all frequencies, with the CNN slightly outperforming the WF.}
    \label{fig:CNN_multi}
\end{figure}

An evaluation on the entire test dataset is shown in Fig.~\ref{fig:CNN_multi}. The top part of the Fig.~\ref{fig:CNN_multi}~(a) shows a 2D histogram of the individual residual, \rind, defined by:

\begin{equation}
    r_{\text{ind}}=\overline{\frac{\abs{\mathrm{ASD}\left(s-\tilde{s}\right)}^2}{\abs{\mathrm{ASD}\left(s\right)}^2}},
    \label{eq:IndividualResidual}
\end{equation}

where $\tilde{s}$ denotes either a network or WF prediction, the overline denotes the average over time  and ASD is the amplitude spectral density. This quantity, in contrast to $\sqrt{R}$ (cf. Eq.~{(\ref{eq:SimulatedResidual})}), is calculated on individual events instead of dataset means. As such, it gives the relative residual of each event after mitigation. A residual of 1 denotes no suppression, while a residual below 1 signifies mitigation. The $x$-axis shows the root mean square (RMS) of the initial signal's ASD; this is used as a measure for the amount of NN contained in the event, or the `signal strength'. The mean residual per RMS-bin for network and WF are added in magenta and yellow, respectively. Error bars show the uncertainty on the mean. The dataset mean of network mitigation and simulation RMS are shown by the horizontal and vertical dashed lines.
The bottom part of the figure shows a histogram of the number of events in each RMS bin in blue, as well as the ratio of the WF and neural network mean per-bin residuals in magenta.

Both WF and CNN show residuals on the single to low double percent level and they decrease for events with higher RMS. The ratio between WF and network mitigation ranges between 1.1 and 1.4 indicating similar performance of both mitigation methods. The network reaches an average residual across the entire dataset of under 5\%.
Fig.~\ref{fig:CNN_small}~(b) shows the ASD of the raw and mitigated signals from WF and CNN over the entire dataset in the top plot, as well as the ratios between the mitigated and raw signal ASDs below. The mitigated noise ASDs of the CNN and WF lie closely together indicating similar mitigation performance. Starting from a flat raw ASD, the mitigated ASDs are also approximately flat. The CNN reduces the noise ASD by a factor 15 to 20, the WF reduction is between 10 and 16. As for the residuals, the WF is outperformed slightly by the network, though both reach similar mitigation factors between 10 and 20.

We conclude, that on this approximately stationary wave field, both WFs and neural networks are able to suppress NN by factors of 10 and above. The network further outperforms the WF reaching mitigation factors up to 20. This performance is relatively constant over frequency, with some losses towards the edges of the examined frequency space. We assume, that these are edge effects and may be reduced by increasing the considered frequency space outside the region of interest.

\subsection{Dominant Wave Events: Plane Waves}

For NN signals, cases of (known or unknown) sources that appear for short or extended periods of time can alter the stationary wave field assumed before. To gain more insight into these cases, we investigate the mitigation performance of neural networks and WFs in the case of a singular dominating noise source.
We model these as time series containing only a single plane wave of fixed frequency and amplitude. We retrain the CNN and WF on events of this kind using the regular 8-seismometer grid and examine their performance.

\begin{figure}
    \makebox[\textwidth]{
        \subfloat[]{\includegraphics[width=0.53\linewidth]{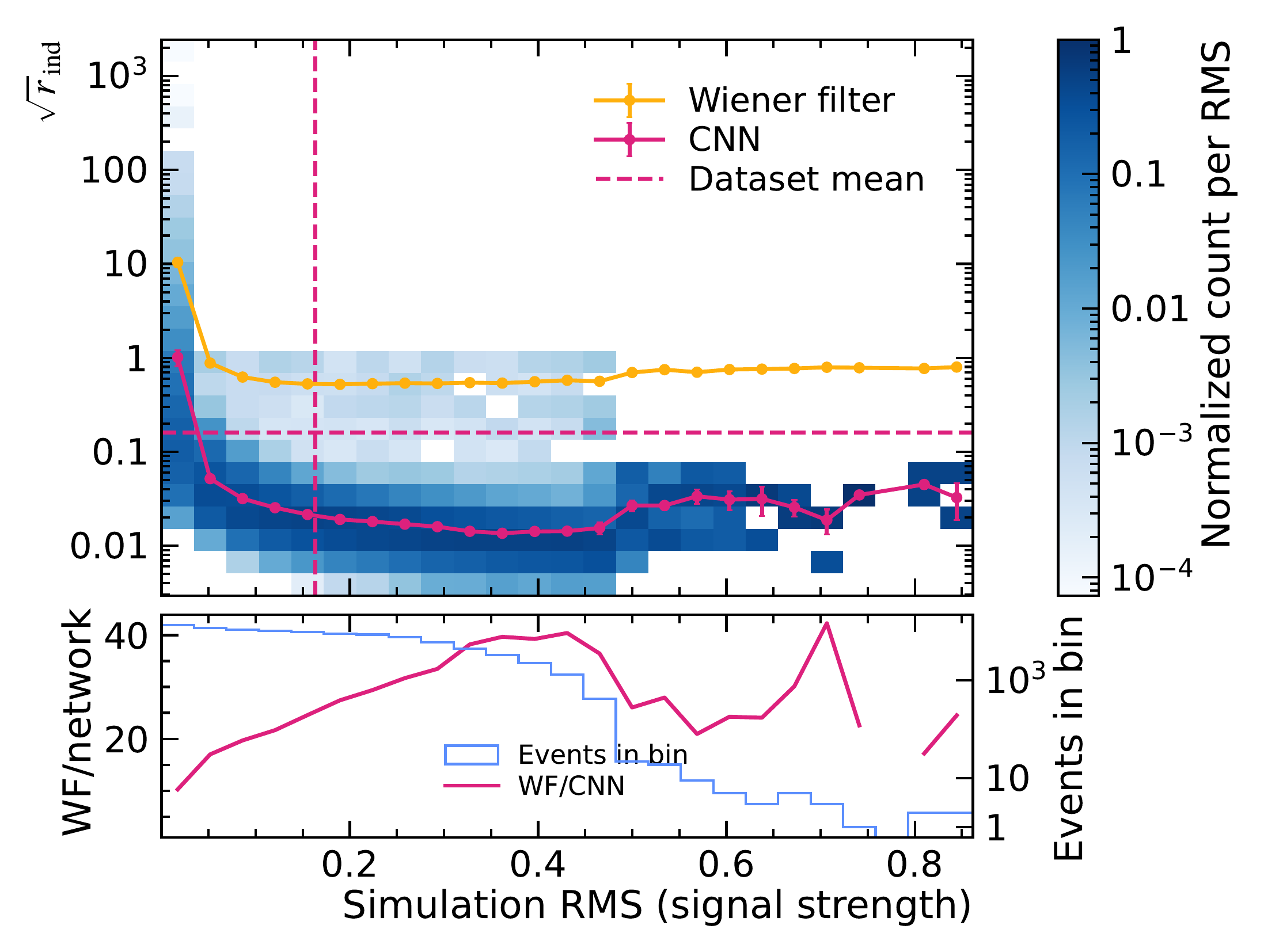}\label{fig:CNN_plain_r}}
        \subfloat[]{\includegraphics[width=0.47\linewidth]{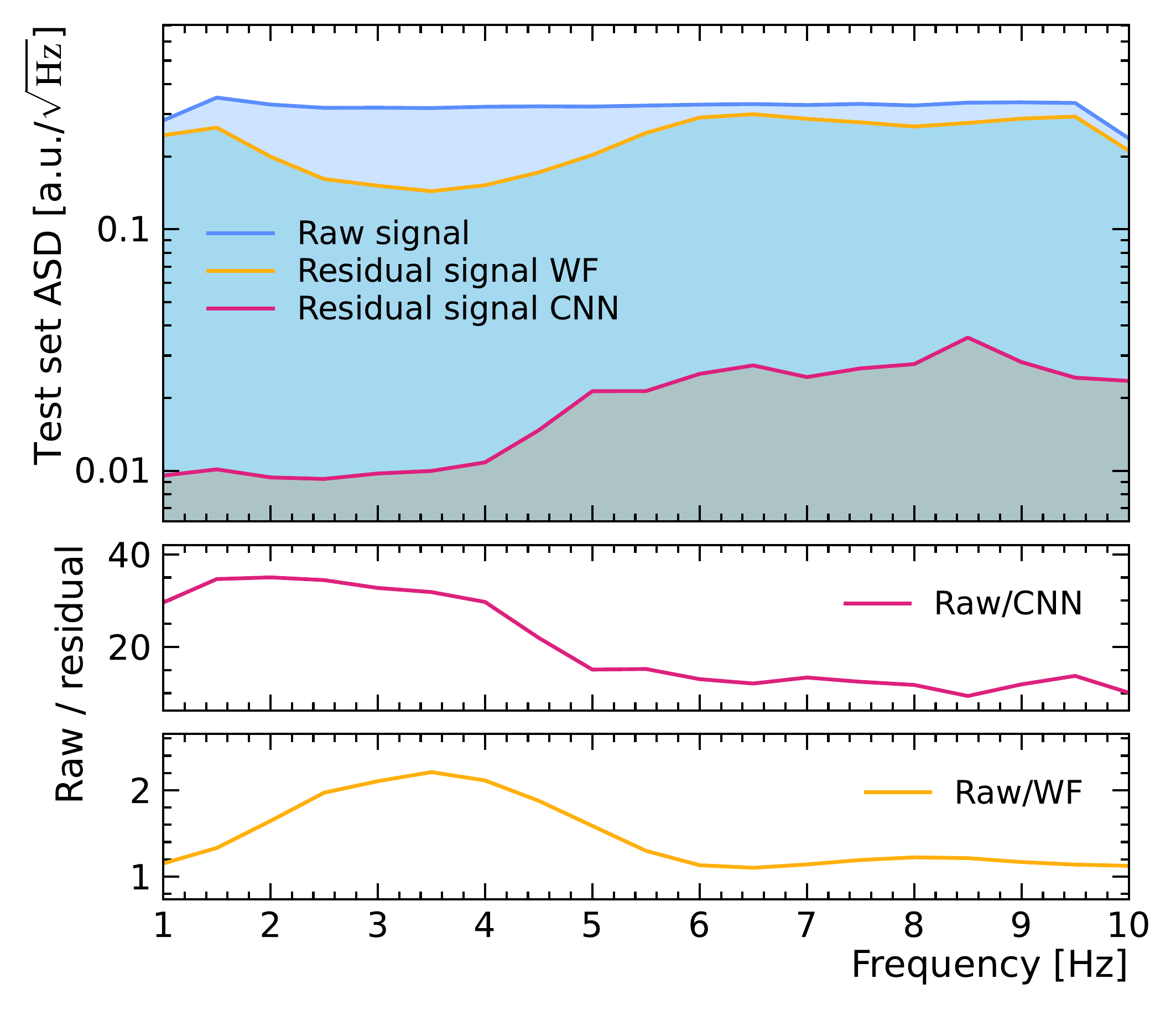}\label{fig:CNN_plain_ASD}}
    }
    \caption{Results of training a CNN on single plane waves using the regular {8-seismometer} array: (a) \rind\ of CNN and WF and the ratio between residuals. Compared to the stationary case, WF residuals increase significantly. CNN residuals are up to 40 times smaller than those of the filter and remain mostly in the single-percent region, below 20\%. (b) ASDs of raw, WF- and CNN-mitigated events. CNNs outperform WFs and have suppression factors between 10 and 40 depending on frequency. The WF does not exceed a factor of 2 in suppression.}
    \label{fig:CNN_plain}
\end{figure}

The \rind\ for this case are shown in Fig.~\ref{fig:CNN_plain}~(a). Similar to the stationary case, the network's residuals remain mostly in the single-percent range, except for the lowest RMS events where they are about an order of magnitude larger. The 
WF residuals are significantly larger. They start at 10 for low-RMS events and approach values slightly below 1. With increasing RMS, residuals increase slightly for the WF, while they drop further for the network. Overall, the network reaches a dataset average residual below 20\% and the ratio between CNN and WF residuals reaches values up to 40.
The raw and mitigated ASDs in Fig.~\ref{fig:CNN_plain}~(b) show a similar picture. The ASD of the WF-mitigated signal is significantly higher than that of the CNN-mitigated signal. The WF only improves the noise by a factor 2, while the CNN can reach mitigation factors between 10 and 40. 

We conclude that when the noise is no longer approximately stationary, neural networks can outperform the WF.
Indeed, the stationarity assumption is not fulfilled once NN is dominated by distinct events. These events can change in amplitude and direction and as such do not exhibit stationary behavior. Even with constant amplitudes, a change of direction alone is enough to result in a non-stationary case. As the direction of the wave changes, the distribution of the signal between the three spatial dimensions (signal channes) changes. Through the suspension, the mirror movement is forced on one selected spatial dimension which also breaks isotropic behaviour in the signal channels. This leads to a change in SNR, which breaks stationarity. Under these circumstances, neural networks can perform significantly better than (non-adaptive) WFs.

\subsection{Dominant Wave Events: Wave Packets}

Besides dominant NN events of long duration, it is also possible, that short burst-like events dominate the noise background. We model these as Gaussian wave packets and evaluate the performance of neural networks and WFs on these events. We also investigate the potential benefits of different model architectures and seismometer array layouts on these transients.

\begin{figure}
    \makebox[\textwidth]{
        \subfloat[]{\includegraphics[width=0.53\linewidth]{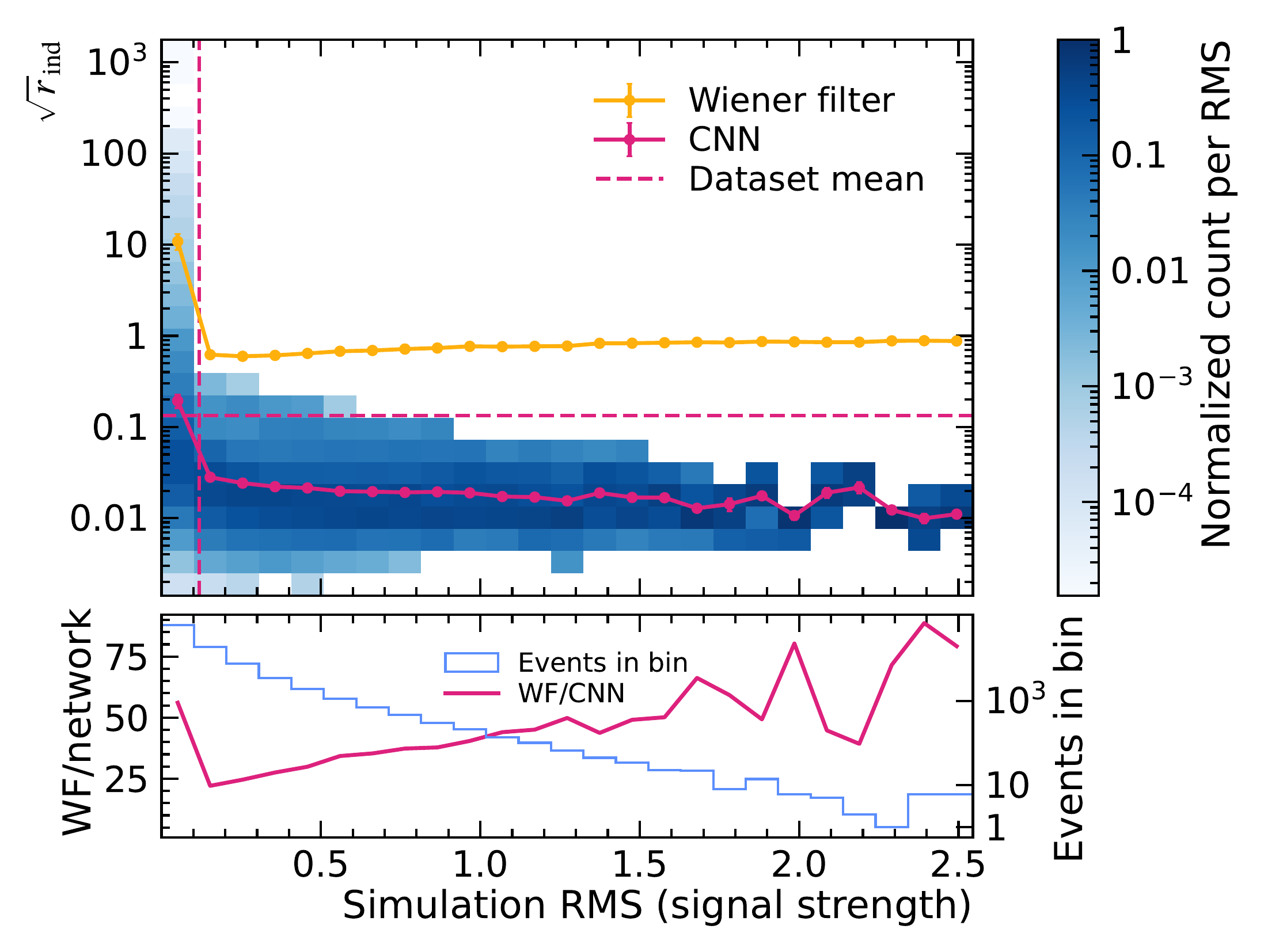}\label{fig:CNN_small_r}}
        \subfloat[]{\includegraphics[width=0.47\linewidth]{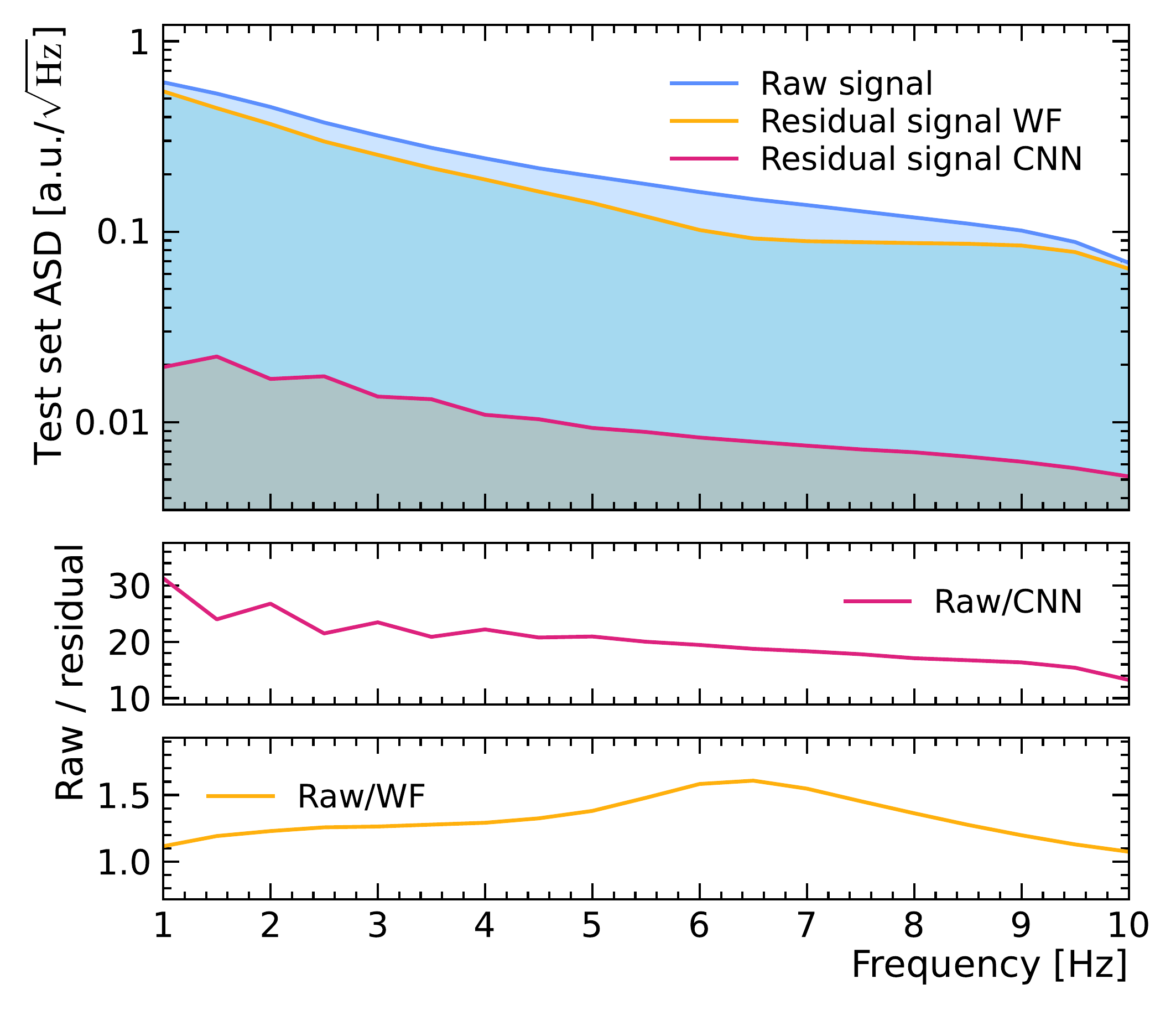}\label{fig:CNN_small_ASD}}
    }
    \caption{Results of training a CNN on Gaussian wave packets using the regular {8-seismometer} array: (a) Network mitigation as a factor of signal strength. The average residual across the entire dataset is $13\%$. The CNN outperforms the WF, especially for strong events, by factors of up to 80. (b) ASDs of raw, WF- and CNN-mitigated events. Again, CNNs outperform WFs and have an average suppression factor of 20, and at least 10 for all frequencies.}
    \label{fig:CNN_small}
\end{figure}

The first case we examine is the CNN trained on the 8 sensor grid. The results are shown in Fig.~\ref{fig:CNN_small}.
As for the single plane wave case, the network and WF residuals are worse for low-RMS events, and drop towards higher RMS. The network's per-bin-means are all below one, reaching a dataset average of 13\%. The mitigation of the WF is significantly worse. The ratio between WF and network residual ranges from \SIrange{25}{80}{}, with the highest values towards high-NN events.

In Fig.~\ref{fig:CNN_small}~(b), the ASDs of the raw and mitigated signals are shown. While the WF does not exceed a  mitigation factor of $1.6$, the CNN reaches an order of magnitude stronger noise suppression across all frequencies. The strongest suppression, of about a factor 30, is achieved for low frequencies. As in the plane wave case, neural networks are able to beat WFs for non-stationary signals. In the following, we investigate the performance under changes to the seismometer array grid and the neural network architecture.

\begin{figure}
    \makebox[\textwidth]{
        \subfloat[]{\includegraphics[width=0.53\linewidth]{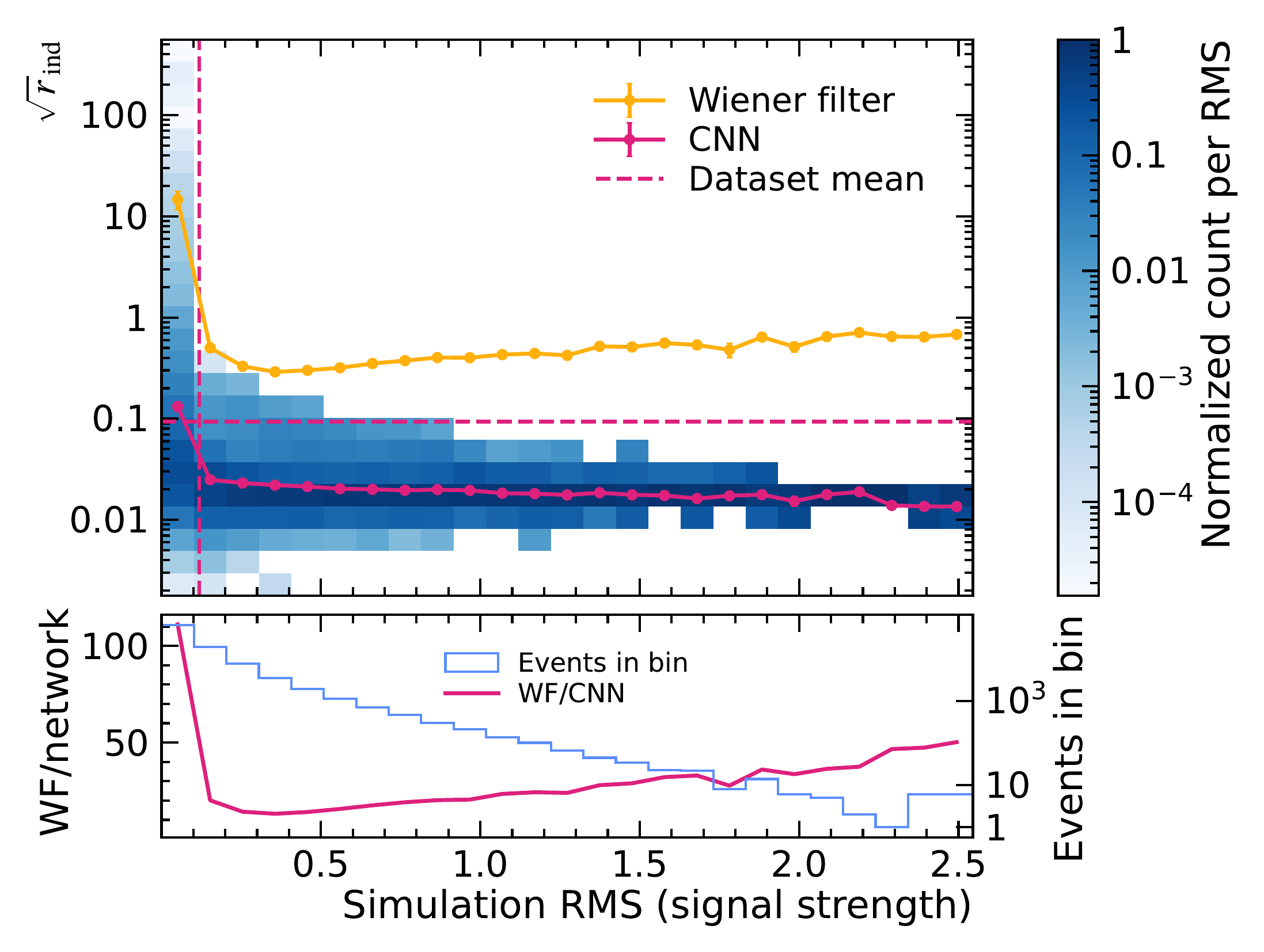}\label{fig:CNN_large_r}}
        \subfloat[]{\includegraphics[width=0.47\linewidth]{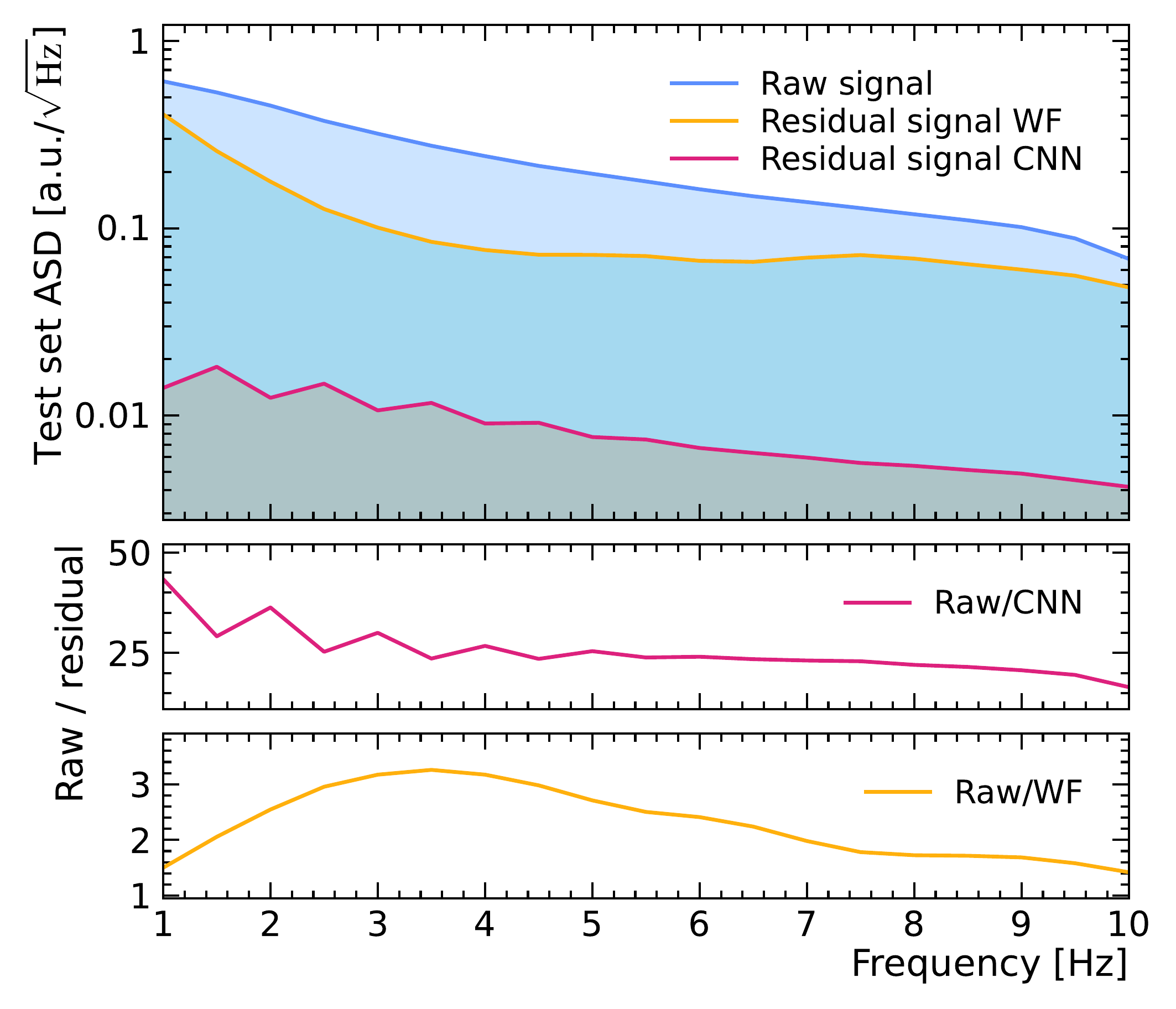}\label{fig:CNN_large_ASD}}
    }
    \caption{CNN trained on gaussian wave packets using the larger seismometer array grid with 32 sensors: (a) \rind\ for the network and the WF and the CNN average \rind\ (top panel), and ratio between per-bin averages of CNN and WF (bottom panel). The network has lower residuals by a factor \SIrange{10}{110}{}. The mean residual of the network decreases to $9\%$. (b) ASD of the raw and mitigated signals, as well as ratios between them. The CNN's average ASD suppression factor is 25.}
    \label{fig:CNN_large}
\end{figure}

The first modification is an extension of the array from 8 to 32 seismometers. Both CNN and WFs were determined and evaluated on the larger array. Fig.~\ref{fig:CNN_large}~(a) shows the residuals, \rind, and the ASDs of the raw and mitigated signals.
The \rind\ of network and WF have improved compared to the 8-seismometer case, with the network now showing an average residual across the dataset of $\EW{\rind}=9\%$. As before, they result in the worst residuals for low-NN events. With the exception of these events, the ratio between WF and network has decreased, indicating that the performance of the WF has benefited more from the larger array than the neural network. In Fig.~\ref{fig:CNN_large}~(b), the ratio between raw and mitigated NN has also increased for both, CNN and WF. The former shows a maximum mitigation factor of almost $50$ for \SI{1}{Hz} and an average of 25. As for \rind, the WF experiences a stronger performance increase than the network, as the suppression factor (yellow curve in the bottom plot of Fig.~\ref{fig:CNN_large}~(b)) increases by up to a factor of $2$ compared to Fig.~\ref{fig:CNN_small}~(b).

While both CNN and WF, experience improvements and the CNN still outperforms the WF, the filter seems to benefit more. We interpret this as the ability of the network to handle smaller arrays better and extraction information more efficiently.

\begin{figure}
    \makebox[\textwidth]{
        \subfloat[]{\includegraphics[width=0.53\linewidth]{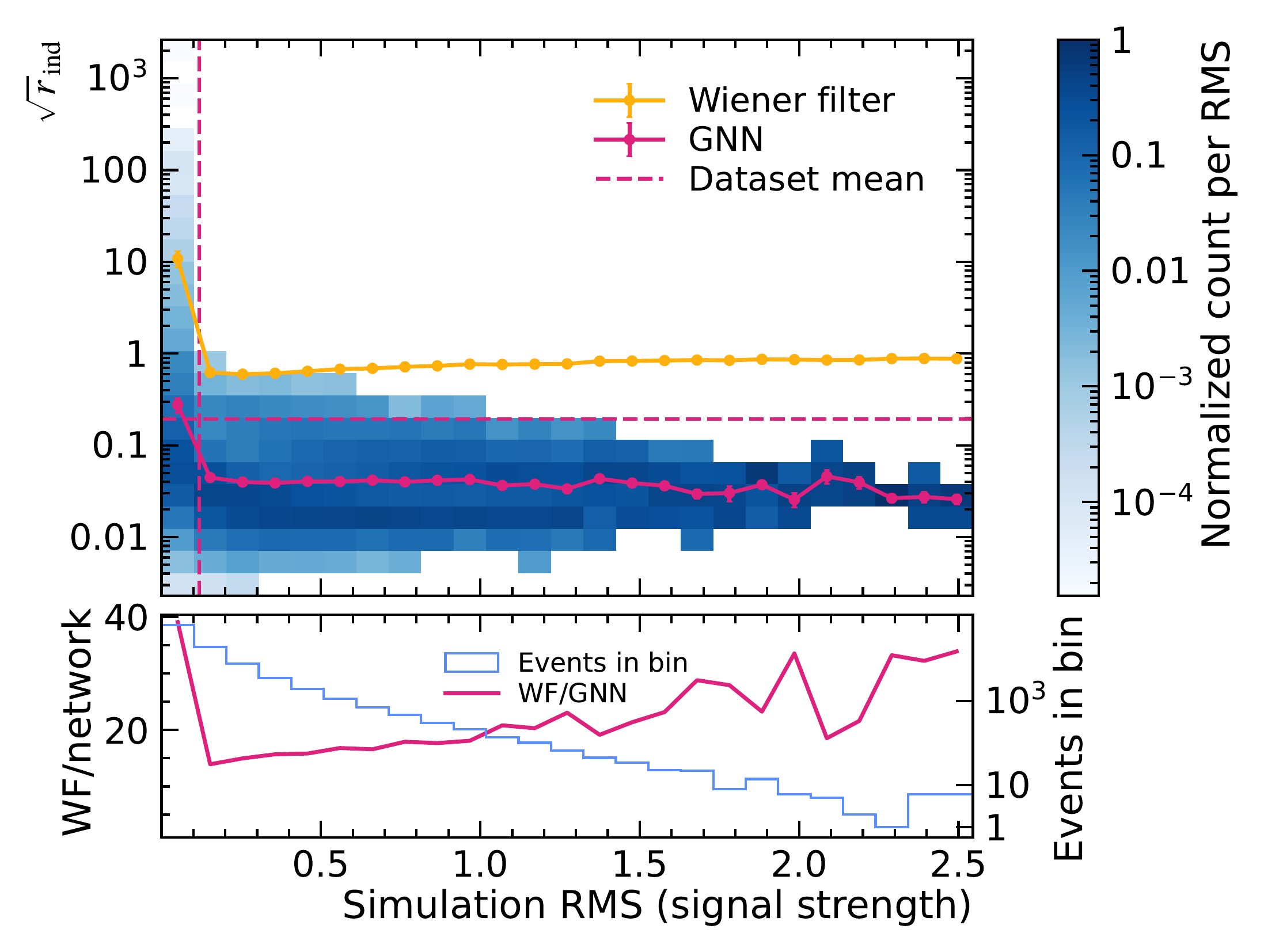}\label{fig:GNN_small_r}}
        \subfloat[]{\includegraphics[width=0.47\linewidth]{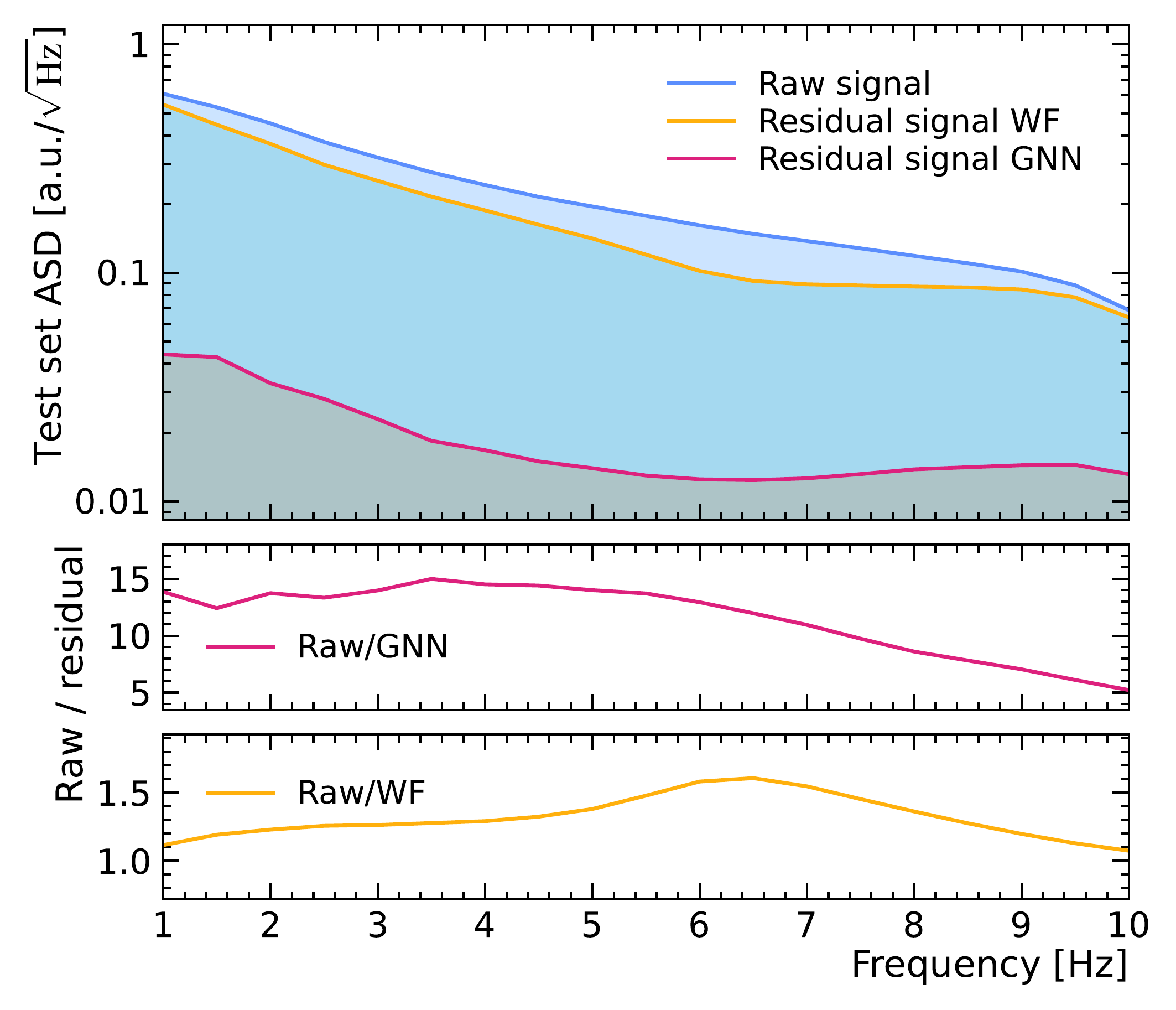}\label{fig:GNN_small_ASD}}
    }
    \caption{GNNs on Gaussian wave packets using the regular {8-seismometer} grid as a benchmark: (a) \rind\ for GNN and WF and the GNN average \rind\ (top panel), as well as the ratio between the per-bin means (bottom panel). The ratio between GNN and WF residuals ranges from \SIrange{15}{40}{}. The GNN reaches an average $\EW{\rind}=19\%$. (b) ASDs of raw and mitigated signals, as well as the ratios between them, with the GNNs reaching ratios as high as 15 compared to the WF's ratio of~$1.6$. The GNN shows lower performance compared to the CNN on this regular grid (cf.~\ref{fig:CNN_small}).}
    \label{fig:GNN_small}
\end{figure}

CNNs can only be deployed on regular grids. For a realistic underground sensor array, this is not a suitable assumption for several reasons: Non-regular grids have proven to be more effective for NN mitigation~\cite{FrancescaSingleMirror, FrancescaJointMirrorOptimization, OurPaper, ArrayOptimizationForVirgo,ophardt2025silencing}.
Furthermore, some borehole locations might be inaccessible due to surface property ownership. We propose GNNs as a solution, which are designed to handle unstructured graphs. Information about the distance between seismometers is provided using the adjacency matrix.

To compare the GNN to the CNN, we train and evaluate a model on the regular 8-seismometer grid as a benchmark. 
In Fig.~\ref{fig:GNN_small}~(a), the dataset mean of \rind\ from the GNN residuals is $19\%$, compared to $13\%$ for the CNN (cf. Fig~\ref{fig:CNN_small}~(a)). Since the WF is the same in both cases, its performance does not change. The ratio between WF and network is smaller than for the CNN but the GNN still outperforms the WF by more than a factor of 10 for all values of RMS.
In Fig.~\ref{fig:GNN_small}~(b), the average ratio between raw and network-mitigated signal
is still as large as 12,
but smaller than the factor of 20 for the CNN (cf. Fig.~\ref{fig:CNN_small}~(b)).

Some performance is lost when switching from CNNs to GNNs. A possible explanation for this decrease is the reduced number of trainable parameters in GNNs for similarly complex architectures. The specific architecture employed possesses only $40\%$ of the trainable parameters compared to the corresponding CNN. Some of the lost performance might be recovered by optimizing and expanding the architecture.

\begin{figure}
    \makebox[\textwidth]{
        \subfloat[]{\includegraphics[width=0.53\linewidth]{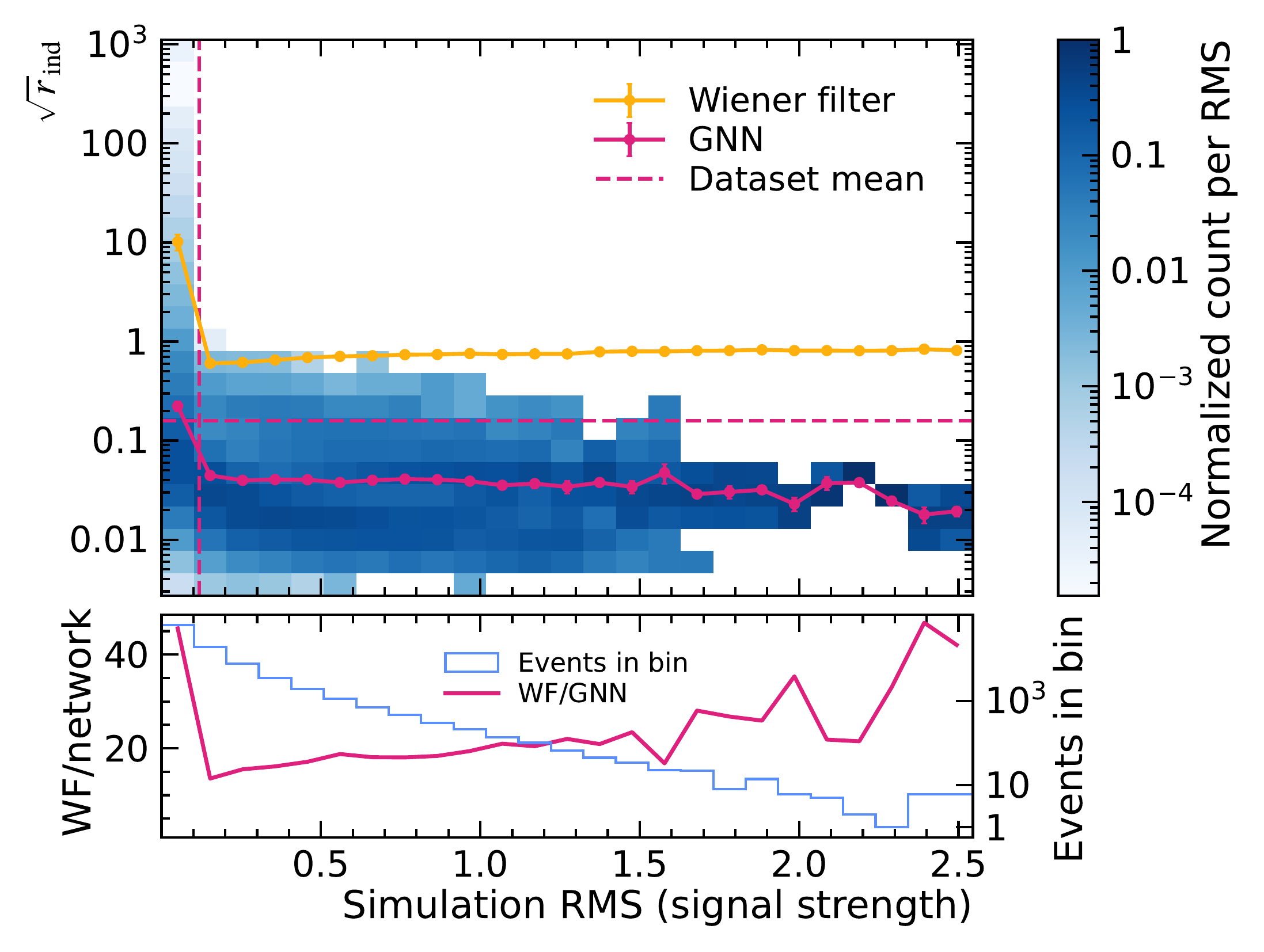}\label{fig:GNN_opt_r}}
        \subfloat[]{\includegraphics[width=0.47\linewidth]{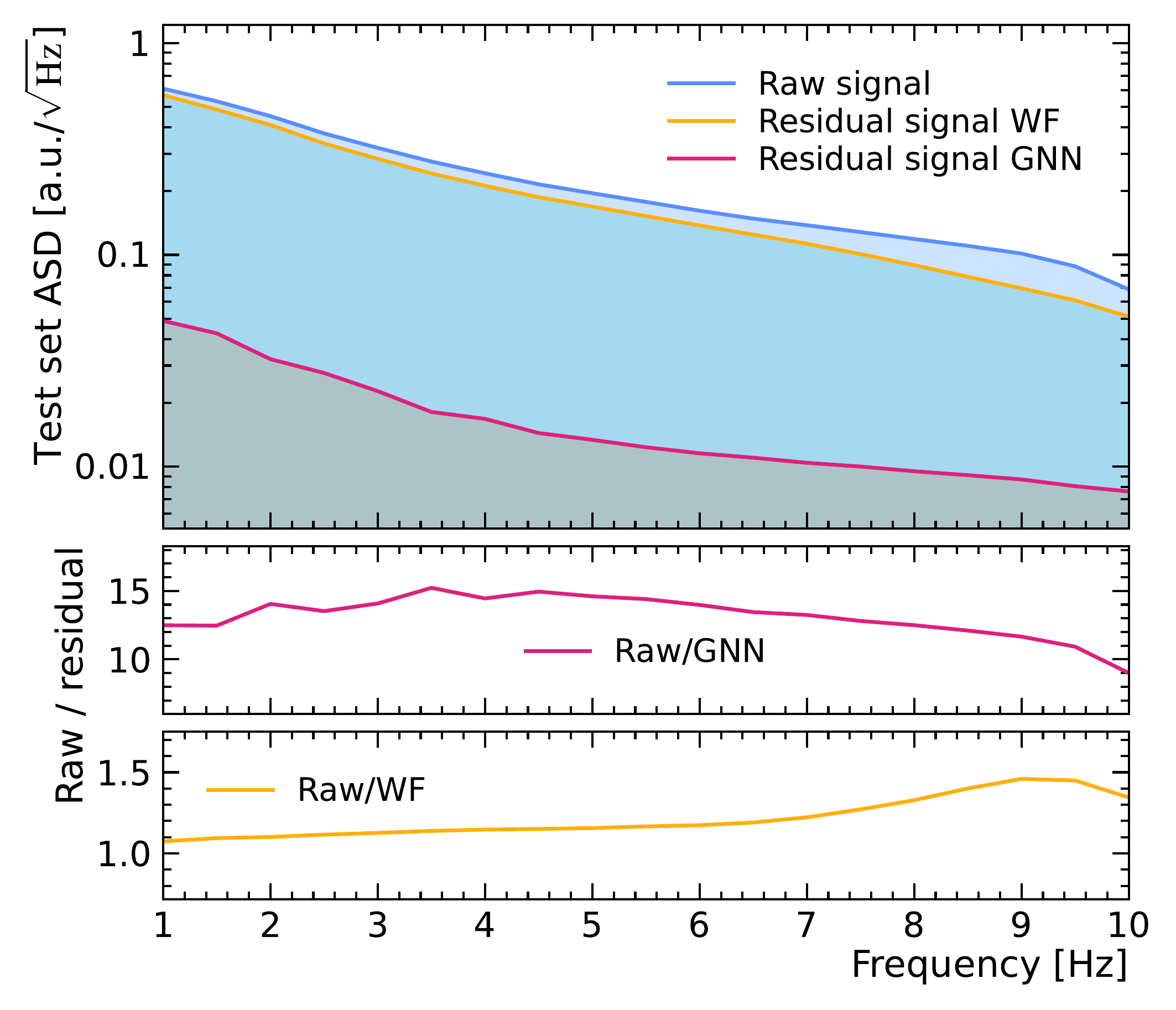}\label{fig:GNN_opt_ASD}}
    }
    \caption{GNN results for Gaussian wave packets using 8 sensors with positions optimized for best NN mitigation on $\SI{10}{Hz}$: (a) \rind\ and dataset average for the GNN and per-bin averages for network and WF (top panel). Across the dataset, the GNN achieves an average residual of $16\%$. The ratio between GNN and WF per-bin averages (bottom panel) ranges from \SIrange{15}{45}{}. (b) Ratio between raw and mitigated ASDs ranges from \SIrange{1}{1.5}{} for the filter and from \SIrange{9}{16}{} for the network.}
    \label{fig:GNN_opt}
\end{figure}

Another possibility to increase performance is to make use of optimized seismometer arrays, deviating from the original grid.
One such grid, consisting of 8 seismometers and optimized for a frequency of $\SI{10}{Hz}$ is studied as an example in the following. More information on the optimization process is given in Ref.~\cite{OurPaper}.

The per-bin mean residuals in Fig.~\ref{fig:GNN_opt_r} for the GNN show residuals between $2\%$ and $30\%$ with a dataset mean of $\EW{\rind}=16\%$. The WF residuals are significantly higher, by factors \SIrange{15}{45}{}. The raw ASD in Fig.~\ref{fig:GNN_opt_ASD} is suppressed by a factor of 10 across almost all frequencies by the GNN, while the WF only reaches suppression factors of no more than $1.5$. Both, network and WF are capable of coping with irregular grid configurations. As before, the network outperforms the WF.

While both, GNNs and WFs, are capable of mitigating NN based on irregular seismometer arrays, these results indicate that GNNs outperform WFs in this task on the Gaussian wave packets.
The benchmark configuration with 8 seismometers that was studied here is only a representative example and different optimization conditions might lead to better results.
In this example, the mitigation potential of the WF is very similar to the potential on the regular grid (cf. Fig.~\ref{fig:GNN_small_r}), while the GNN performance slightly improved, comparing Fig.~\ref{fig:GNN_opt_ASD} and Fig.~\ref{fig:GNN_small_ASD}.
Although there may certainly be additional gains in both cases from array optimization, these results suggest that GNNs have a stronger NN mitigation potential than WFs on irregular grids on transient NN.

\section{Conclusions}
\label{sec:conclusions}
In this paper, we have introduced a Newtonian-noise simulation and investigated the mitigation capabilities of neural networks and compared them to the Wiener filter.
The simulation generates time series of a displacement field and the corresponding Newtonian-noise force in an infinite homogeneous medium. Using this, we generated datasets based on approximately stationary, plane-wave-like and Gaussian-wave-packet-like density fluctuations for several seismometer array grid configurations.
We trained neural networks on these datasets to predict the Newtonian noise from the seismometer data. The network architecture was based on flexible spatiotemporal convolutions to examine both convolutional and graph attention networks. We then compared the mitigation of neural networks with the Wiener filter.

We have found that neural networks represent not only a suitable alternative, but may provide considerable performance improvements, depending on the nature of the seismic wave field. In all cases of single, dominating noise sources that we studied, we found neural networks to mitigate NN better than Wiener filters. On approximately stationary wave fields, we found them to match their performance, independent of signal strength or frequency. In all other cases studied, we found neural networks to suppress the NN spectrum by more than an order of magnitude on average.

We further examined the benefits of different array configurations and model architectures on transient-like noise scenarios. Although graph attention networks cannot yet match the performance of convolutional neural networks, they bridge the problem of regular grids for CNNs. In the future we assume that at least a part of the performance gap can be closed by increasing the amount of trainable parameters. Increasing the size of the seismometer array improves the results of networks and Wiener filters, as expected.
The networks were found to extract information more effectively from smaller arrays. This could potentially decrease the amount of boreholes needed, cutting costs. Optimizing the position of the seismometers may further enhance NN mitigation performance.

The above results show that neural networks are able to mitigate NN effectively and have significant performance advantages over Wiener filters in scenarios where the seismic wave field is not stationary.
This indicates that neural networks can play an important role in achieving low-frequency sensitivity goals even at higher surrounding noise levels.
In the future, it would be interesting to compare their performance to adaptive Wiener filters, as studied in Ref.~\cite{VirgoNewtonianNoiseCancellingSystemPart2}.
Future research directions should also target more realistic seismic and geological models, including layered or fractured underground geometries, surface waves, reflections, and seismic mode conversions, with the goal of studying NN mitigation in realistic seismic wave fields, as well as tests with data from subsurface borehole measurements, following ideas from Ref.~\cite{Koley:2023okf}.

\section*{Acknowledgments}
This research was supported by the German Federal Ministry of Research, Technology and Space (BMFTR) via project 05A2023 under grant number 05A23PA1.
We thank Martin Erdmann for providing computing resources on the VISPA cluster.

\printbibliography

\end{document}